\documentclass[a4paper,fleqn]{cas-sc}

\usepackage{lineno,hyperref}
\usepackage{bm}
\usepackage{subcaption}
\usepackage[numbers]{natbib}

\begin{document}

\shorttitle{Immersed boundary methods and geometric singularities}
\shortauthors{}
\title [mode = title]{Handling geometric singularities efficiently with the immersed-boundary method: application to riblets}

\author[1]{Stefano Cipelli}[orcid=0009-0001-7228-5549]
\ead{stefano.cipelli@kit.edu}
\credit{Methodology, Investigation, Validation, Formal analysis, Writing - Original Draft}
\cormark[1]

\author[2]{Alessandro Chiarini}[orcid=0000-0001-7746-2850]
\ead{alessandro.chiarini@polimi.it}
\credit{Methodology, Formal analysis, Writing - Review \& Editing}

\author[2]{Maurizio Quadrio}[orcid=0000-0002-7662-3576]
\ead{maurizio.quadrio@polimi.it}
\credit{Methodology, Writing - Review \& Editing, Project administration}

\author[1]{Davide Gatti}[orcid=0000-0002-8178-9626]
\ead{davide.gatti@kit.edu}
\credit{Methodology, Investigation, Validation, Formal analysis, Writing - Review \& Editing, Supervision}

\author[3]{Paolo Luchini}[orcid=0000-0001-6527-7762]
\ead{luchini@cplcode.net}
\credit{Conceptualization, Methodology, Software, Writing - Review \& Editing, Supervision}

\affiliation[1]{
    organization={Institute of Fluid Mechanics, Karlsruhe Institute of Technology},
    addressline={Kaiserstraße 10},
    city={Karlsruhe},
    postcode={76131},
    country={Germany}
    }
\affiliation[2]{
    organization={Dept. of Aerospace Engineering, Politecnico di Milano},
    addressline={Campus Bovisa, via La Masa 34},
    city={Milano},
    postcode={20156},
    country={Italy}
    }
\affiliation[3]{
    organization={Dip. di Ingegneria Industriale, Universit\`a di Salerno},
    addressline={via Giovanni Paolo II 132},
    city={Fisciano},
    postcode={84084},
    country={Italy}
    }

\cortext[1]{Corresponding author}


\begin{abstract}
We present an enhancement of classical immersed boundary methods for the direct numerical simulation of the incompressible Navier--Stokes equations, designed to efficiently cope with geometrical singularities. The method, while general, is described here in the context of the flow over a plane wall covered by small longitudinal riblets. 
The enhancement consists in prescribing that the solution near the boundary matches the steady Stokes solution. In this way, sharp corners, where velocity gradients are strongest and viscous effects dominate, become tractable with grids comparable to those necessary for a smooth wall. 
The method requires the availability of a local Stokes solution, that can be obtained either analytically or numerically. Two test cases are discussed: the laminar flow over one riblet, which allows to measure protrusion heights, and a fully turbulent channel flow with ribbed walls. 
In both cases, the method reproduces reference results at a substantially lower computational cost.
\end{abstract}

\begin{highlights}
\item Efficient immersed-boundary treatment of sharp corners via local Stokes-based correction.
\item Accurate riblet-tip resolution without body-fitted grid refinement.
\item Validation on laminar and turbulent riblet flows against reference DNS data.
\item Accurate prediction of protrusion heights and drag reduction at reduced cost.
\end{highlights}


\maketitle

\section{Introduction}
\label{section:intro}
Direct Numerical Simulation (DNS) of the Navier--Stokes equations requires resolving all the dynamically significant spatial and temporal scales of the flow. 
In canonical wall-bounded turbulent flows, the most stringent requirement is typically set by the smallest dissipative scales, whose characteristic size is comparable to the Kolmogorov length, and reaches its minimum in the viscous layer. This scale becomes progressively smaller at higher Reynolds numbers \citep{pope-2000}. Since, at the same time, the largest energy-containing motions must be captured using sufficiently large computational domains to ensure statistical convergence \citep{lozano-duran-jimenez-2014a}, DNS of high-Reynolds turbulent flows is one of the most demanding challenges in computational physics.

Such large resolution requirements, however, may no longer suffice when the flow domain presents geometric singularities like corners or edges. 
Sharp geometrical details cause unbounded velocity gradients, so that some local quantities do not converge under mesh refinement \citep{pathria-karniadakis-1995,moffatt-2019}; even quantities that remain finite are highly sensitive to the near-corner resolution and require a very fine clustering of grid points near the singularity, with a corresponding increase in computational cost. 
Flows over geometries with sharp features are relatively common, and include flows over backward-facing steps \citep{kaiktsis-karniadakis-orszag-1996, le-moin-kim-1997}, ducts and channels with non-smooth cross-sections \citep{gavrilakis-1992, pinelli-etal-2010, marin-etal-2016, orlandi-modesti-pirozzoli-2018}, walls with mounted obstacles \citep{yakhot-etal-2006, lee-sung-krogstad-2011, diaz-daniel-laizet-vassilicos-2017} and flows around bluff bodies with corners \citep{tamura-miyagi-kitagishi-1998, bruno-etal-2010, chiarini-quadrio-2022}.

An important case, of specific interest for the present paper, is the flow over a flat wall covered by drag-reducing riblets. Riblets are small streamwise-aligned grooves, forming an anisotropic roughness capable of reducing turbulent skin-friction drag under suitable conditions \citep{walsh-1980, bechert-etal-1997}. Some of their key performance indicators, e.g. protrusion heights, are defined in the laminar regime \cite{luchini-manzo-pozzi-1991}. 
Despite decades of experimental and numerical investigations, riblets remain an active research topic. Recent studies have addressed their combined effect on momentum and heat transfer \cite{stalio-nobile-2003,rouhi-etal-2022}, the role of secondary flows and Kelvin--Helmholtz instabilities in the drag-reduction breakdown \cite{endrikat-etal-2021a,modesti-etal-2021,camobreco-etal-2025}, their complex effect on drag in the drag-increasing regime \cite{gatti-etal-2020a,deyn-gatti-frohnapfel-2022,endrikat-etal-2022}, three-dimensional riblet geometries \cite{cafiero-amico-iuso-2024}, and riblet modelling \cite{wong-etal-2024,luchini-chung-2026}, to name a few.

DNS remains essential for a complete understanding of riblets effects. Two main strategies have so far been employed for the DNS of turbulent flows over ribbed walls. The first, and most natural, involves the use of body-fitted grids with strong grid refinement near the most severe geometrical singularities --- the riblet tips --- to represent the ribbed walls. This strategy has been adopted since the earliest DNS studies dating back to the 1990s, such as those by \citet{choi-moin-kim-1993} and \citet{chu-karniadakis-1993}, who simulated the flow over V-shaped riblets in turbulent channels using second-order finite differences and spectral elements, respectively. Later, body-conforming meshes were also adopted by Stalio and Nobile \citep{stalio-nobile-2003} and by a number of more recent studies \citep{endrikat-etal-2021a,endrikat-etal-2021,modesti-etal-2021,endrikat-etal-2022,rouhi-etal-2022} with second-order finite volumes and spectral element methods \cite{chan-etal-2023}. 
The main drawback of this approach is that the riblet-tip singularity is not treated explicitly, except through strong local grid refinement. As a result, the local order of the underlying numerical methods is reduced \citep{pathria-karniadakis-1995}, and the computational cost increases significantly. The resulting extreme cost associated with resolving the riblet tips is typically mitigated by reducing the Reynolds number \citep{choi-moin-kim-1993,chu-karniadakis-1993,stalio-nobile-2003}, by adding riblets to only one wall of a turbulent channel while measuring the reference friction on the opposite wall \citep{choi-moin-kim-1993,chu-karniadakis-1993,stalio-nobile-2003}, and by adopting computational domains of limited streamwise and spanwise extent. The latter mitigation has been used, to different degrees, in all the studies mentioned above, partially compromising the representation of the larger scales of motion.

A second strategy is to avoid body-conforming meshes, by representing the ribbed walls via an immersed boundary method (IBM). The first riblet simulation of this kind was performed by \citet{goldstein-handler-sirovich-1995}, who --- also in later works \citep{goldstein-tuan-1998,strand-goldstein-2011} --- introduced V-shaped riblets via a force-field-based IBM \citep{goldstein-handler-sirovich-1993} into a classical spectral discretisation of turbulent channel flow \cite{kim-moin-moser-1987}. These studies were mainly focused on particular riblet-related phenomena, such as the dynamics of turbulent spots over riblet walls \citep{strand-goldstein-2011} or riblet-induced secondary flows \citep{goldstein-tuan-1998}, and were not aimed at accurately reproducing riblet geometries or quantitatively measuring their drag-reduction performance. Similarly, \citet{garcia-mayoral-jimenez-2011} improved the earlier IBM method of \citet{goldstein-handler-sirovich-1995} to achieve quantitative accuracy for the specific case of rectangular riblets. For this geometry, a Cartesian grid of collocation points can be arranged to coincide with the riblet contours, and the geometric singularity is comparatively weak. Their method combines Fourier series in the wall-parallel directions with second-order finite differences in the wall-normal direction, using a collocated discretisation enabled by approximate mass conservation \citep{nordstrom-mattsson-swanson-2007} together with the IBM of \citet{iaccarino-verzicco-2003}. Turbulent channel flows over riblets with rectangular cross-sections have also been simulated via IBM by \citet{yoon-el-samni-chun-2006} and \citet{sasamori-etal-2017}, the latter featuring three-dimensional riblets that are not aligned with the mean flow direction.

The IBM potentially offers a cost-effective way to represent complex solid boundaries in existing numerical solvers originally developed for canonical turbulent flows. Although instrumental in providing the first numerical evidence of drag reduction by riblets, the IBM formulations discussed above do not explicitly address the geometric singularity at the riblet tips. As a result, accurate simulations still require extremely fine grid resolutions in the vicinity of the tips, which offsets the potential computational advantages of the IBM. Consequently, these approaches remain limited to small Reynolds numbers, limited domain sizes and --- at least for those achieving quantitative accuracy \citep{yoon-el-samni-chun-2006,garcia-mayoral-jimenez-2011,sasamori-etal-2017} --- to rectangular riblet geometries, whose $90^\circ$ angles imply a comparatively weak singularity.
 
Properly representing the sharp tips of riblets is essential. Riblets are effective only if their tips are sufficiently sharp: rounded or blunted edges reduce their drag-reducing capability significantly, as quantified by \cite{garcia-mayoral-jimenez-2011a}. Additionally, in simulations, insufficient resolution leads to inaccurate drag predictions. 
Adequate resolution of the flow near edges and corners can be achieved through aggressive local mesh refinement, albeit with a dramatic increase of the computational cost: the number of degrees of freedom grows, and severe timestep restrictions due to stability constraints slow down the simulation. 
However, the singular nature of the flow near sharp corners can be exploited to achieve accuracy at lower cost. In the immediate vicinity of a corner, viscous effects dominate over inertial and unsteady terms, owing to the large velocity gradients. As a result, the local flow is well approximated by the steady Stokes solution for the same geometry.
For spectral and high-order methods applied to lid-driven cavity flows, for instance, \citet{botella-peyret-1998, auteri-quartapelle-vigevano-2002} improved accuracy by subtracting analytically derived contributions to account for corner singularities, while \citet{schultz-lee-boyd-1989} demonstrated that a Chebyshev pseudospectral method achieves rapid convergence when the leading singular behaviour at corners is removed analytically. \citet{pathria-karniadakis-1995} showed that the accuracy of pseudospectral methods can be restored if a mapping and base transformation is adopted to account for a corner singularity. 
Others have used matched asymptotic expansions or combined analytical/numerical schemes, in which a local Stokes solution near the corner is coupled with a global numerical solution elsewhere \citep[e.g.,][]{shi-breuer-durst-2004}. 
For finite-difference discretization, \citet{luchini-1991} introduced a method that locally modifies the Navier--Stokes equations near geometric singularities by adding deferred correction terms to enforce the proper Stokes-like behaviour. 
The guiding principle is always to incorporate the known singular behaviour into the numerical formulation to preserve accuracy without requiring a prohibitive local refinement.

The present work describes how to enable an immersed-boundary method to effectively cope with sharp geometrical features. 
Despite the aforementioned advantages of IBM in terms of simplicity and computational efficiency, their applicability to flows with sharp corners has so far been hampered by the tip-singularity problem outlined above. Here, we address these challenges by extending the corner-correction approach of \citet{luchini-1991} to riblet-covered walls.
Our method combines the simplicity and efficiency of a second-order finite-difference scheme on a Cartesian grid with the high-fidelity treatment of geometric singularities. The underlying numerical method is the discrete-forcing immersed-boundary formulation recently introduced by \citet{luchini-etal-2025}; a deferred correction is added to enforce the steady Stokes solution in the region around sharp corners. Thus, instead of imposing a linear velocity profile between the wall and the first fluid node, the solver prescribes the correct Stokes behaviour, significantly improving near-corner accuracy without additional grid refinement. Since the correction is applied only to a small set of grid points near the singularity and relies on precomputed coefficients, the computational overhead is negligible.
The paper is organised as follows. Section~\ref{section:method} summarises the DNS solver and the baseline IBM formulation. Section~\ref{section:coco} introduces the corner-correction methodology, including the derivation of the correction terms and the computation of the local Stokes solutions. In Section~\ref{section:results}, the method is validated against a Stokes-flow benchmark and applied to turbulent channel flows with riblets, with comparisons against reference DNS performed on body-fitted grids. Finally, conclusions and perspectives are discussed in Section~\ref{section:conclusions}.

\section{DNS solver and immersed-boundary method}
\label{section:method}

The present work builds on the DNS code introduced by \citet{luchini-2016}, which solves the incompressible Navier–Stokes equations in primitive variables on a staggered Cartesian grid for channel-flow configurations. Solid boundaries are treated using the IBM introduced by \citet{luchini-etal-2025}, enabling the representation of complex geometries. The corner-correction (COCO) proposed in this paper replaces the immersed-boundary treatment in the vicinity of sharp corners.
Second-order finite differences are employed in all spatial directions. The Cartesian coordinates $x$, $y$, and $z$ denote the streamwise, spanwise, and wall-normal directions, respectively, with corresponding velocity components $u$, $v$, and $w$. The momentum equations are advanced in time using a fractional-step method combined with a third-order Runge–Kutta (RK3) scheme \cite{rai-moin-1991}, with the time step adjusted to maintain a constant CFL number.
The time advancement is explicit, except for the treatment of the IBM/COCO forcing terms. 
The Poisson equation for pressure is solved by an iterative successive over-relaxation (known as SOR) algorithm applied to a Gauss-Seidel red-black iteration. 
Algorithms based on low-order finite-differences and the use of an IBM are now standard in the context of DNS of turbulent channel flows \cite{choi-moin-kim-1993, orlandi-2006, costa-2018}.
The solver applies periodic boundary conditions in the streamwise and spanwise directions. 
For simplicity the following discussion assumes constant grid spacings ($\Delta x$, $\Delta y$ and $\Delta z$) in all directions.
The remainder of this section presents the IBM introduced by \citet{luchini-etal-2025}, by reformulating it as a second-order deferred correction to the staircase approximation of the Navier--Stokes equations in the vicinity of a solid wall.
In the following, $u_\alpha(i,j,k)$ denotes the $\alpha$-th discretized velocity component at the grid point $(i,j,k)$. 
For continuous fields, the notation $u_\alpha(x,y,z)$ is used.

The momentum equation for the $\alpha$-th velocity component can be written as
\begin{linenomath}
\begin{equation} \label{eq:ns}
\frac{\partial u_\alpha}{\partial t}
= - \frac{\partial u_\alpha u_\beta}{\partial x_\beta} 
- \frac{1}{\rho} \frac{\partial p}{\partial x_\alpha}
+ \nu \frac{\partial^2 u_\alpha}{\partial x_\beta \partial x_\beta}
- \nu \mathcal{F}_\alpha ,
\end{equation}
\end{linenomath}
where $\rho$ is the fluid density, $\nu$ its kinematic viscosity, and $\mathcal{F}_\alpha$ denotes the correction/forcing term of the IBM.
The discretization at grid points adjacent to the boundary is identical to that employed throughout the rest of the computational domain. In absence of $\mathcal{F}_\alpha$, this results in a first-order staircase representation of the boundary.
The enhancement of the approximation order is achieved through the $\mathcal{F}_\alpha$ term, which enforces a linear velocity profile in the vicinity of the solid boundary, thereby recovering the second-order accuracy of the underlying spatial discretization.
The deferred correction term is evaluated based on the local velocity value and is expressed as $\mathcal{F}_\alpha = \lambda_\alpha^{\mathcal{I}} u_\alpha$, where $\lambda_\alpha^{\mathcal{I}}$ is the IBM coefficient.

When time-discretizing Equation~\eqref{eq:ns} using a third-order Runge--Kutta scheme, treating the entire right-hand side explicitly recovers a canonical explicit, extrapolation-based IBM.
To avoid the associated poor stability properties, the IBM term is treated implicitly.
Although an exact integration of the $\mathcal{F}_\alpha$ term is possible while retaining the second-order temporal accuracy of the scheme \citep{luchini-etal-2025}, the IBM forcing term is here integrated using an implicit Euler scheme for simplicity.
The time-discrete momentum equation for the $\alpha$-th velocity component, at a generic time step $n$ and at the $k$-th Runge--Kutta stage, can then be written as follows, with the nonlinear, viscous, and pressure-gradient contributions grouped into the term $f_\alpha$:
\begin{linenomath}
\begin{equation} \label{eq:rk}
u_\alpha^{*,n+\frac{k}{3}} =
\frac{1}{1+\Delta t  \nu \lambda^{\mathcal{I}} \left( a_k + b_k \right)}
\left[
u_\alpha^{n+\frac{k-1}{3}}
+ \Delta t \left(
a_k f_\alpha^{n+\frac{k-1}{3}}
+ b_k f_\alpha^{n+\frac{k-2}{3}}
\right)
\right],
\end{equation}
\end{linenomath}
where $u_\alpha^*$ denotes the intermediate velocity.
According to \citet{rai-moin-1991}, the coefficients $a_k$ and $b_k$ appearing in Equation~\eqref{eq:rk} are given by
\begin{linenomath}
\begin{equation} \label{eq:ab}
\begin{aligned}
a_k &= \left\{ \frac{64}{120}, \frac{50}{120}, \frac{90}{120} \right\}, \quad &
b_k &= \left\{ 0, -\frac{34}{120}, -\frac{50}{120} \right\} \, .
\end{aligned}
\end{equation}
\end{linenomath}

\begin{figure}
    \centering
    \begin{subfigure}[b]{0.38\textwidth}
        \centering
        \includegraphics[width=\linewidth]{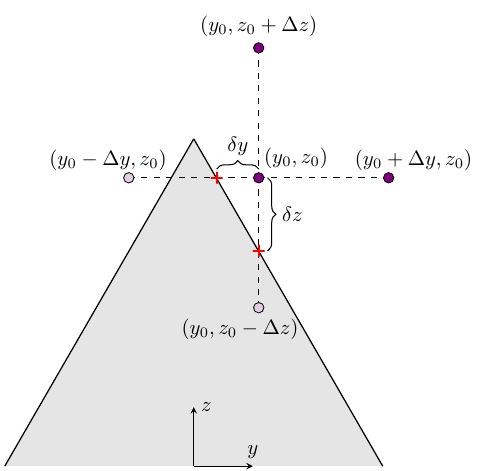}
        \caption{}
        \label{fig:stencil_ibm}
    \end{subfigure}
    \hspace{1.5cm}
    \begin{subfigure}[b]{0.38\textwidth}
        \centering
        \includegraphics[width=\linewidth]{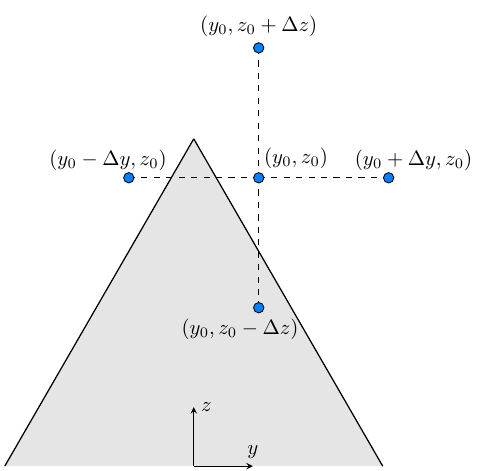}
        \caption{}
        \label{fig:stencil_coco}
    \end{subfigure}
    \caption{A visual representation of the numerical stencil for the classical immersed-boundary implementation (a) and for the corner correction (b). The IBM makes use of the the distances from the central point to the stencil intersections with the boundary ($\delta x_{l, m/p}$) to calculate the coefficients. Corner correction, on the other hand, always uses the Laplace/Stokes solutions at the stencil points, despite their position around the tip and intersections with the boundary.}
    \label{fig:stencil}
\end{figure}

The IBM considered here corrects the viscous Laplacian term, which dominates near solid walls where velocity gradients are largest.
The correction is illustrated using a two-dimensional Laplace equation for a scalar field $u(y,z)$; see Figure~\ref{fig:stencil_ibm}.
Here, $y$ and $z$ denote the transverse and vertical directions, respectively.
We consider the grid point $(j,k)$, with neighboring points $(j-1,k)$ and $(j,k-1)$ located inside the solid body.
The discretized Laplace equation at $(j,k)$, corresponding to the physical location $(y_0,z_0)$, including the deferred-correction term, is then written as
\begin{linenomath}
\begin{equation} \label{eq:lapl-ibm}
d^{(2)}_y(-1)   \underbrace{u(j-1,k)}_{0}
+ d^{(2)}_z(-1) \underbrace{u(j,k-1)}_{0}
+ \left[ d^{(2)}_y(0) + d^{(2)}_z(0) \right]  u(j,k)
+ d^{(2)}_y(+1)  u(j+1,k)
+ d^{(2)}_z(+1)  u(j,k+1)
- \mathcal{F}(j,k)
= 0 .
\end{equation}
\end{linenomath}
The coefficients $d^{(2)}_y(\cdot)$ and $d^{(2)}_z(\cdot)$ represent the Laplacian discretization coefficients associated with the present spatial discretization at all stencil points ($-1/+1$: stencil points in the negative/positive $y$ or $z$ direction; $0$: central stencil point).
The terms $u(j-1,k)$ and $u(j,k-1)$ vanish, as the $(j-1,k)$ and $(j,k-1)$ grid points lie within the solid phase.
According to the configuration shown in Figure~\ref{fig:stencil_ibm}, the term $\mathcal{F}(j,k)$ is employed to impose a linear velocity profile in the near-wall region, calibrated on the value of $u(j,k)$.
Specifically, $\mathcal{F}(j,k)$ is determined such that the field $u_e(y,z)$,
\begin{linenomath}
\begin{equation} \label{eq:linear}
u_e(y,z)
= \left(
1 + \frac{y - y_0}{\delta y}
+ \frac{z - z_0}{\delta z}
\right) u(j,k) ,
\end{equation}
\end{linenomath}
is a solution of Equation~\eqref{eq:lapl-ibm}.
Here, $\delta y$ and $\delta z$ denote the distances from the boundary in the negative $y$ and $z$ directions, respectively, while the coordinates $(y_0, z_0)$ identify the location of the $(j,k)$ grid point.
By substituting Equation~\eqref{eq:linear} into Equation~\eqref{eq:lapl-ibm}, the following expression is obtained:
\begin{linenomath}
\begin{equation} \label{eq:deferred-ibm}
\mathcal{F}(j,k)
=
\underbrace{
\left[
d^{(2)}_{y}(0)
+ d^{(2)}_{z}(0)
+ \left( \frac{\delta y + \Delta y}{\delta y} \right) d^{(2)}_{y}(+1)
+ \left( \frac{\delta z + \Delta z}{\delta z} \right) d^{(2)}_{z}(+1)
\right]
}_{\lambda^{\mathcal{I}}(j,k)}
u(j,k).
\end{equation}
\end{linenomath}
The factor enclosed in square brackets defines the IBM coefficient $\lambda^{\mathcal{I}}(j,k)$ at the $(j,k)$ grid point. 
This expression can be generalized to the three-dimensional case, allowing for an arbitrary number of stencil intersections with the solid boundary; for the $(i,j,k)$ grid point, one obtains
\begin{linenomath}
\begin{equation} \label{eq:ibm}
\lambda^{\mathcal{I}}(i,j,k)
=
\sum_{l=1}^{3}
\left[
d^{(2)}_{x_l}(-1) \frac{\Delta x_l - \delta x_{l,m}}{\delta x_{l,m}}
+
d^{(2)}_{x_l}(+1) \frac{\Delta x_l - \delta x_{l,p}}{\delta x_{l,p}}
\right].
\end{equation}
\end{linenomath}
Here, $\Delta x_l$ denotes the grid spacing in the $x_l$-th direction, while $\delta x_{l,m}$ and $\delta x_{l,p}$ represent the distances of the $i,j,k$ grid point from the boundary in the negative ($m$) and positive ($p$) $l$-th directions, respectively. Here, we have used the identity $\sum_{i=-1}^{1} d^{(2)}_{x_l}(i) u_e(\bm{x} + i \Delta \bm{x}_l) = 0$, with $\Delta \bm{x}_l = \Delta x_l \hat{\bm{e}}_l$ and $\hat{\bm{e}}_l$ being the unit vector in the $l$ direction, since Equation~\eqref{eq:linear} is an exact solution of the discretized equations, which are based on polynomial interpolation.

If no intersection with the boundary occurs along a given stencil arm, $\delta x_{l,m/p}$ is set equal to $\Delta x_l$.
As a grid point approaches the boundary, the corresponding $\lambda^{\mathcal{I}}$ tends to infinity, driving the velocity in Equation~\eqref{eq:rk} to zero, as expected for a point lying on the solid boundary.
In the absence of any intersection with the boundary, the governing equation remains unmodified by the IBM.
Although the deferred correction operates in the same manner for all velocity components, the coefficients $\lambda^{\mathcal{I}}$ must be computed and stored separately for each component due to grid staggering.

\section{The corner correction}
\label{section:coco}

When considering flows in the vicinity of solid boundaries, where velocity gradients are large, unsteady and convective effects are of lower order compared to viscous effects.
In turbulent channel flows, the region closest to the wall, known as the viscous sublayer, is characterized by an approximately linear mean velocity profile, with the instantaneous velocity field well approximated by the steady Stokes solution.
This rationale is further reinforced in the presence of irregular walls featuring sharp corners, such as those at riblet tips.
These corners, which constitute geometrical singularities, induce even stronger local gradients and, in principle, require infinitely fine mesh resolution to be fully resolved.
Consequently, previous DNS studies of riblet flows have relied on extremely fine, body-fitted grids to accurately capture the flow around sharp corners, resulting in a substantial increase in computational cost \citep{choi-moin-kim-1993, stalio-nobile-2003, el-samni-chun-yoon-2007, garcia-mayoral-jimenez-2011, sasamori-etal-2017, endrikat-etal-2021}.

In this section, we describe an enhancement of the previously introduced IBM implementation based on the steady Stokes solution.
The proposed approach improves the accuracy of the IBM in the vicinity of sharp corners and allows for an accurate discretization of the geometry while employing relatively coarse grids.
The fundamental difference with respect to the original IBM formulation presented in \citet{luchini-etal-2025} lies in the near-wall representation of the solution, which is no longer assumed to vary linearly with the wall distance, but is instead given by the Stokes solution corresponding to the same geometry.

For a given angle of the corner tip, the steady Stokes problem can be solved analytically by enforcing no-slip and no-penetration conditions on the solid boundary and idealizing the tip geometry as an infinite two-dimensional wedge.
In the limiting case of a smooth flat wall, this formulation recovers the classical Couette solution, with a velocity profile that varies linearly with the distance from the wall, and the COCO forcing reduces to the standard IBM forcing.
For riblet geometries, the Stokes solution in the neighbourhood of a sharp corner is derived analytically and used to compute the COCO coefficients for each velocity component.

Considering the geometric configuration shown in Figure~\ref{fig:stencil_coco}, a deferred-correction strategy based on the Stokes solution is adopted. Following the approach described in Section~\ref{section:method}, the discretized Stokes equations are written at the $(j,k)$ grid point with the inclusion of a deferred-correction term. For brevity, only the $v$ velocity component is reported, since the extension to the other components is straightforward. The resulting discrete equation reads
\begin{linenomath}
\begin{equation} \label{eq:Stk-coco}
\begin{gathered}
d^{(2)}_y(-1) v(j-1,k)
+ d^{(2)}_z(-1) v(j,k-1)
+ \left[ d^{(2)}_y(0) + d^{(2)}_z(0) \right] v(j,k)
+ d^{(2)}_y(+1) v(j+1,k)
+ d^{(2)}_z(+1) v(j,k+1) \\
- \frac{1}{\mu} \left( d^{(1)}_y(1) p(j+1,k) + d^{(1)}_y(0) p(j,k) \right)
- \mathcal{F}_y^{\mathcal{S}}(j,k)
= 0 .
\end{gathered}
\end{equation}
\end{linenomath}
Similarly to the IBM formulation, the term $\mathcal{F}_y^{\mathcal{S}}$ enforces the analytically known Stokes solution on an edge $\{\bm{u}^{\mathcal{S}},p^{\mathcal{S}}\}$, scaled by the local value of $v(j,k)$. The imposed velocity and pressure fields are thus defined as
\begin{equation}
  v_e(y,z) = \frac{ v^{\mathcal{S}}(y,z) }{ v^{\mathcal{S}}(y_0,z_0) }  v(j,k),
  \qquad
  p_e(y,z) = \frac{ p^{\mathcal{S}}(y,z) }{ v^{\mathcal{S}}(y_0,z_0) }  v(j,k).
\end{equation}
By imposing that $\{ v_e, p_e \}$ satisfy Equation~\eqref{eq:Stk-coco}, the deferred-correction term takes the form
\begin{linenomath}
\begin{equation} \label{eq:defcor-coco}
\begin{gathered}
  \mathcal{F}_y^{\mathcal{S}}(j,k) =
  \Bigg[
  d_y^{(2)}(0) + d_z^{(2)}(0)
  + \frac{ v^{\mathcal{S}}(y_0+\Delta y,z_0) }{ v^{\mathcal{S}}(y_0,z_0) } d_y^{(2)}(+1)
  + \frac{ v^{\mathcal{S}}(y_0,z_0+\Delta z) }{ v^{\mathcal{S}}(y_0,z_0) } d_z^{(2)}(+1) \\
  + \frac{ v^{\mathcal{S}}(y_0-\Delta y,z_0) }{ v^{\mathcal{S}}(y_0,z_0) } d_y^{(2)}(+1)
  + \frac{ v^{\mathcal{S}}(y_0,z_0-\Delta z) }{ v^{\mathcal{S}}(y_0,z_0) } d_z^{(2)}(+1) \\
  - \frac{1}{\mu} \left(
      d_y^{(1)}(1) \frac{ p^{\mathcal{S}}(y_0+\Delta y/2,z_0) }{ v^{\mathcal{S}}(y_0,z_0) }
    + d_y^{(1)}(0) \frac{ p^{\mathcal{S}}(y_0-\Delta y/2,z_0) }{ v^{\mathcal{S}}(y_0,z_0) }
    \right)
  \Bigg] v(j,k) .
\end{gathered}
\end{equation}
\end{linenomath}
The quantity enclosed in square brackets defines the COCO coefficient $\lambda^{\mathcal{S}}(j,k)$ at the grid point $(j,k)$. Unlike the above IBM formulation, this correction can be applied at grid points not immediately adjacent to the riblet wall.

As in the previous section, this formulation naturally extends to three dimensions. For a generic velocity component $u_\alpha$, the COCO coefficient $\lambda^{\mathcal{S}}_\alpha$ at the grid point $(i,j,k)$, corresponding to the physical location $\bm{x}_0=(x_0,y_0,z_0)$, is given by
\begin{equation} \label{eq:lambda-coco}
\begin{gathered}
  \lambda^{\mathcal{S}}_\alpha =
   \sum_{l=1}^3 \left[
      d_l^{(2)}( 0) +
      d_l^{(2)}(-1) \frac{ u_\alpha^{\mathcal{S}}(\bm{x}_0 - \Delta \bm{x}_l ) }{ u_\alpha^{\mathcal{S}}(\bm{x}_0) }
    + d_l^{(2)}(+1) \frac{ u_\alpha^{\mathcal{S}}(\bm{x}_0 + \Delta \bm{x}_l ) }{ u_\alpha^{\mathcal{S}}(\bm{x}_0) }
   \right] \\
   - \frac{1}{\mu} \left(
      d_\alpha^{(1)}(1) \frac{ p^{\mathcal{S}}(\bm{x}_0 + \Delta \bm{x}_\alpha/2 ) }{ u_\alpha^{\mathcal{S}}(\bm{x}_0) }
    + d_\alpha^{(1)}(0) \frac{ p^{\mathcal{S}}(\bm{x}_0 - \Delta \bm{x}_\alpha/2 ) }{ u_\alpha^{\mathcal{S}}(\bm{x}_0) }
   \right),
\end{gathered}
\end{equation}
The above expression applies to all velocity components along the $x$ (longitudinal), $y$ (horizontal), and $z$ (vertical) directions.

Equation~\eqref{eq:lambda-coco} can also be derived through a direct manipulation of the Stokes and momentum conservation equations for a generic velocity component.
As a representative example, we consider the two-dimensional configuration shown in Figure~\ref{fig:stencil_coco} and focus on the $v$ velocity component at the $(j,k)$ grid point, corresponding to the physical location $(y_0,z_0)$. The momentum equation is first rewritten by adding and subtracting the term
$\mu \left[ d_y^{(2)}(0) + d_z^{(2)}(0) \right] v(j,k)$.
By collecting all remaining contributions on the right-hand side into the forcing term $f(j,k)$, the equation can be written as
\begin{linenomath}
\begin{equation} \label{eq:ns-mod}
\frac{\mathrm{d} v(j,k)}{\mathrm{d} t}
=
\mu \left[ d_y^{(2)}(0) + d_z^{(2)}(0) \right] v(j,k)
-
\mu \left[ d_y^{(2)}(0) + d_z^{(2)}(0) \right] v(j,k)
+
f(j,k).
\end{equation}
\end{linenomath}
In the vicinity of sharp corners, the velocity is required to satisfy the Stokes equations. The Stokes equation for the $v$ component is discretized using central finite differences. After rearranging the resulting expression and pre-multiplying by the factor $v(j,k)/v^{\mathcal{S}}(y_0,z_0)$, we obtain
\begin{linenomath}
\begin{equation} \label{eq:stokes-mod}
\begin{aligned}
\left[ d_y^{(2)}(0) + d_z^{(2)}(0) \right] v(j,k)
&=
\Bigg[
\frac{1}{\mu}
\left(
d_y^{(1)}(1) p^{\mathcal{S}}(y_0+\Delta y/2,z_0)
+
d_y^{(1)}(0) p^{\mathcal{S}}(y_0-\Delta y/2,z_0)
\right)
\\
&\quad
-
d_y^{(2)}(-1) v^{\mathcal{S}}(y_0-\Delta y,z_0)
-
d_z^{(2)}(-1) v^{\mathcal{S}}(y_0,z_0-\Delta z)
\\
&\quad
-
d_y^{(2)}(+1) v^{\mathcal{S}}(y_0+\Delta y,z_0)
-
d_z^{(2)}(+1) v^{\mathcal{S}}(y_0,z_0+\Delta z)
\Bigg]
\frac{ v(j,k) }{ v^{\mathcal{S}}(y_0,z_0) } .
\end{aligned}
\end{equation}
\end{linenomath}
Combining Equations~\eqref{eq:ns-mod} and \eqref{eq:stokes-mod} then yields
$ \text{d} v(j,k)/\text{d} t = f(j,k) - \nu \mathcal{F}_y^{\mathcal{S}}(j,k)$
where $\mathcal{F}_y^{\mathcal{S}}(j,k) = \lambda_y^{\mathcal{S}}(j,k) v(j,k)$ coincides with the definition given in Equation~\eqref{eq:defcor-coco}.

This derivation also clarifies why the correction can be applied uniformly at all grid points, including those not immediately adjacent to the wall, without corrupting the Navier--Stokes solution away from corners. The corrected momentum equation can be interpreted as
\begin{linenomath}
\begin{equation} \label{eq:coco-interp}
\frac{\mathrm{d} v(j,k)}{\mathrm{d} t}
= f(j,k)
+ \underbrace{\mathcal{L}^{\mathcal{S}}(v_e)}_{=\,0}
- \mathcal{L}_\Delta^{\mathcal{S}}(v_e) ,
\end{equation}
\end{linenomath}
where $f(j,k)$ is the full discretized Navier--Stokes right-hand side, $\mathcal{L}^{\mathcal{S}}$ denotes the continuous Stokes operator (Laplacian plus pressure gradient), $\mathcal{L}_\Delta^{\mathcal{S}}$ its finite-difference discretization, and $v_e$ is the rescaled Stokes solution defined above. Because $v_e$ satisfies the continuous Stokes equations exactly, $\mathcal{L}^{\mathcal{S}}(v_e) = 0$, and the correction reduces to $-\mathcal{L}_\Delta^{\mathcal{S}}(v_e)$, i.e.\ the term $-\nu \lambda_y^{\mathcal{S}} v(j,k)$ appearing in the equations.
Near a sharp corner, viscous effects dominate and the discretized Navier--Stokes operator is well approximated by $\mathcal{L}_\Delta^{\mathcal{S}}$; the correction then forces the solution to satisfy the Stokes equations exactly rather than only through their (inaccurate, near the singularity) discretization.
Far from the corner, the Stokes field is smooth, so the finite-difference discretization converges to the continuous operator, $\mathcal{L}_\Delta^{\mathcal{S}}(v_e) \to \mathcal{L}^{\mathcal{S}}(v_e) = 0$, and the correction vanishes, leaving the Navier--Stokes equations unaltered. This is also the reason why the correction involves precisely the Stokes operator applied to the local solution, rather than, for example, the Laplacian alone or the full Navier--Stokes operator: the Stokes operator is the only one that $v_e$ satisfies exactly, which guarantees the correct limiting behaviour both near and far from the corner.

Therefore, the evaluation of the deferred-correction term $\bm{\mathcal{F}}^{\mathcal{S}}$ requires knowledge of the local Stokes solution $\{\bm{u}^{\mathcal{S}}, p^{\mathcal{S}}\}$.  
In the following sections, we present the analytical solution of the Stokes problem specifically tailored to the riblet configuration.  
Since the streamwise direction is homogeneous, the pressure-gradient term vanishes in the longitudinal momentum equation, reducing the Stokes problem to a Laplace equation for the corresponding velocity component.

\subsection{Stokes solution for the longitudinal velocity component} 
\label{section:coco-u}

We begin by considering the streamwise velocity component, which is decoupled from the transverse $y$-$z$ plane components.  
Based on the hypotheses outlined above, in the immediate vicinity of geometrical singularities the streamwise velocity should satisfy the Stokes equation
$ \nabla^2 u - \mu^{-1} \partial_x p = 0$.
The boundary conditions are the standard no-slip and no-penetration conditions at the wall.  
Due to the homogeneity of the geometry along the $x$-direction, the pressure-gradient term $\partial_x p$ vanishes, reducing the equation to the Laplace equation $\nabla^2 u = 0$. Similarly, the pressure-gradient term disappears from the corresponding Equation~\eqref{eq:lambda-coco}.

To solve the Laplace equation, we consider a two-dimensional infinite wedge with the tip angle and orientation of interest. 
This allows the Laplace solution $u^{\mathcal{S}}$ to be obtained in closed form as a combination of eigensolutions. 
Since the COCO correction is applied only in the immediate vicinity of the corners, this approach can be used for an array of finite-size riblets; within this neighborhood, the Laplace solution is the same in both configurations.

The expression for $u^{\mathcal{S}}$, the solution of the Laplace equation, can be written in polar coordinates $\{r, \theta\}$ centered at the riblet tip (see Figure~\ref{fig:stokes}) as
\begin{linenomath}
\begin{equation} \label{eq:u-lapl}
u^{\mathcal{S}}(r,\theta) = A r^{m} \cos\left( m \theta \right), \quad m = \frac{\pi}{2 \varphi_w},
\end{equation}
\end{linenomath}
where $\varphi_w = \pi - \alpha/2$, with $\alpha$ denoting the riblet tip angle, and $\theta$ measured from the riblet symmetry axis.  
The value of $m$ is chosen as the smallest among the possible eigenvalues, representing the dominant mode of the present problem, since the solution is intended to be valid for vanishing $r$.  
$A$ is a constant whose exact value is irrelevant, as the COCO coefficient $\lambda_{u_j}^{\mathcal{S}}$ only involves ratios of $u^{\mathcal{S}}$ values (see Equation~\eqref{eq:lambda-coco}).  
For illustration, the $u^{\mathcal{S}}$ field is plotted in Figure~\ref{fig:analyticalstokes} around a wedge geometry with a $60^\circ$ tip angle.

\subsection{Stokes solution for pressure and the transverse and vertical velocity components}
\label{section:coco-vw}

We now turn to the $v^{\mathcal{S}}$ and $w^{\mathcal{S}}$ velocity components in the transverse plane, i.e., the components along the non-homogeneous directions.  
Unlike $u^{\mathcal{S}}$, the pressure $p^{\mathcal{S}}$ does not vanish in this case, and the full Stokes equations must be satisfied.  
As in the Laplace problem, the no-slip and no-penetration boundary conditions are imposed at the wall.  
Following the approach outlined in \S\ref{section:coco-u}, the geometry is modeled as an infinite wedge, which allows for the formulation of an analytical eigensolution for both velocity components and the pressure field.

\begin{figure}
    \centering
    \begin{subfigure}[t]{0.47\textwidth}
        \centering
        \includegraphics[width=0.8\linewidth]{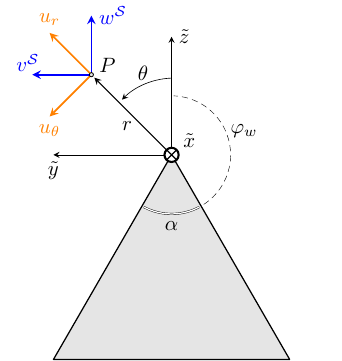}
        \caption{}
        \label{fig:stokes}
    \end{subfigure}
    \hfill
    \begin{subfigure}[t]{0.47\textwidth}
        \centering
        \includegraphics[width=0.8\linewidth]{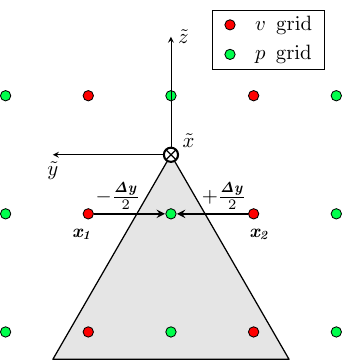}
        \caption{}
        \label{fig:polydromy}
    \end{subfigure}
\caption{(a) Reference system for the Stokes problem. $(r,\theta)$ are the point coordinates w.r.t. the riblet tip. $\varphi_w$ is defined as $\varphi_w = \pi - \alpha/2$, where $\alpha$ is the riblet tip angle. The relation linking $(u_r,u_\theta)$ to $(v^{\mathcal{S}},w^{\mathcal{S}})$ is reported in Equation~\eqref{eq:stokes12}. \hspace{19mm} (b) Pressure and spanwise velocity staggered grids. When the pressure gradient term needed in Euqation~\eqref{eq:Stk-coco} requires a pressure point sitting outside the fluid domain, the choice is to interpolate the value for pressure between the two riblet sides, so that the function for the artificial pressure inside the solid body is continuous and single-valued.}
\label{fig:cocofig}
\end{figure}

The Stokes eigensolutions can be expressed in polar coordinates $(r,\theta)$ centered at the riblet tip (see Figure~\ref{fig:stokes} for the reference system).  
Let $u_r$ and $u_\theta$ denote the in-plane velocity components in the radial and tangential directions, respectively.  
The vorticity component in the $x$-direction, $\omega$, and the streamfunction, $\psi$, are defined as
\[
\omega = \frac{\partial (r u_\theta)}{\partial r} - \frac{\partial u_r}{\partial \theta}, 
\quad
u_r = \frac{1}{r} \frac{\partial \psi}{\partial \theta}, 
\quad
u_\theta = - \frac{\partial \psi}{\partial r}.
\]  
With these definitions, the Stokes equations can be rewritten in terms of $\omega$ and $\psi$ as
\begin{linenomath}
\begin{align} \label{eq:stokes1}
\nabla^2 \omega &= 0, &
\nabla^2 \psi &= - \omega.
\end{align}
\end{linenomath}
This system of PDEs can be simplified by separation of variables:
\begin{linenomath}
\begin{align} \label{eq:stokes2}
\omega(r,\theta) &= R(r) G(\theta), &
\psi(r,\theta) &= P(r) F(\theta) .
\end{align}
\end{linenomath}
Substituting these forms into Equation~\eqref{eq:stokes1} and requiring the solutions to be regular near the corner, one obtains
\begin{linenomath}
\begin{align} \label{eq:stokes3}
R(r) &= r^{\gamma-1}, &
P(r) &= r^{\gamma+1},
\end{align}
\end{linenomath}
where $\gamma$ is still unknown.
The Stokes problem is thus reduced to a pair of ordinary differential equations in $\theta$:
\begin{linenomath}
\begin{align} \label{eq:stokes4}
G''(\theta) + (\gamma-1)^2 G(\theta) &= 0, &
F''(\theta) + (\gamma+1)^2 F(\theta) &= G(\theta) .
\end{align}
\end{linenomath}
Based on Equations~\eqref{eq:stokes1} and \eqref{eq:stokes3}, the solutions to these ODEs take the form
\begin{linenomath}
\begin{equation} \label{eq:stokes5}
\begin{aligned}
G(\theta) &= A_1 \cos\big((\gamma-1) \theta\big) + A_2 \sin\big((\gamma-1) \theta\big), \\
F(\theta) &= B_1 \cos\big((\gamma+1) \theta\big) + B_2 \sin\big((\gamma+1) \theta\big)
           + B_3 \cos\big((\gamma-1) \theta\big) + B_4 \sin\big((\gamma-1) \theta\big) .
\end{aligned}
\end{equation}
\end{linenomath}
Integration constants can be determined by applying the no-slip and no-penetration boundary conditions at the riblet walls ($\theta = \pm \varphi_w$). 
Note that these conditions are imposed on the velocity components only, not on the streamfunction itself; hence, the system is singular and an arbitrary integration constant remains. 
The boundary conditions are:
\begin{linenomath}
\begin{align} \label{eq:stokes6}
\frac{dF}{d\theta}\biggr|_{\theta = \varphi_w} = 0, \quad
\frac{dF}{d\theta}\biggr|_{\theta = -\varphi_w} = 0, \quad
F(\varphi_w) = 0, \quad
F(-\varphi_w) = 0.
\end{align}
\end{linenomath}
These conditions can be rewritten as a linear system for the unknowns $\gamma$ and $\bm{\mathit{B}} = (B_1, B_2, B_3, B_4)$:
\begin{linenomath}
\begin{equation} \label{eq:stokes7}
M(\gamma) \bm{\mathit{B}} = \bm{0},
\end{equation}
\end{linenomath}
where $\gamma$ is obtained by enforcing $\det\big(M(\gamma)\big) = 0$, which leads to the transcendental relation
\begin{linenomath}
\begin{equation} \label{eq:stokes8}
\gamma^2 \sin^2(2 \varphi_w) - \sin^2(2 \gamma \varphi_w) = 0 .
\end{equation}
\end{linenomath}
Solving Equation~\eqref{eq:stokes8} numerically for $\varphi_w \in [\pi/2, \pi]$ yields two acceptable solutions. 
As an example, for riblets with a $60^\circ$ tip angle we obtain:
\begin{linenomath}
\begin{align} \label{eq:stokes9}
\gamma_1 \approx 0.5122214 \quad \text{and} \quad \gamma_2 \approx 0.7309007.
\end{align}
\end{linenomath}
The physical meaning of these two solutions will be discussed later in this section. 
Once a value for $\gamma$ is chosen, the linear system can be solved. 
Since $\det\big(M(\gamma)\big) = 0$, the system is underdetermined; we set $B_1 = 1$ to determine the remaining coefficients. 
All variables below depend only on the riblet tip angle $\varphi_w$. 
For brevity, we define
\[
a_1 = \cos\big((\gamma+1)\varphi_w\big), \quad
a_2 = \sin\big((\gamma+1)\varphi_w\big), \quad
a_3 = \cos\big((\gamma-1)\varphi_w\big), \quad
a_4 = \sin\big((\gamma-1)\varphi_w\big).
\]
Then, the coefficients are
\begin{linenomath}
\begin{equation} \label{eq:stokes10}
\begin{aligned}
B_1 &= 1, & B_2 &= -\frac{a_4}{a_3} \frac{a_2 a_3 (\gamma+1) - a_1 a_4 (\gamma-1)}{a_1 a_4 (\gamma+1) - a_2 a_3 (\gamma-1)},\\[2mm]
B_3 &= - \frac{a_1}{a_3}, & B_4 &= \frac{a_2}{a_3} \frac{a_2 a_3 (\gamma+1) - a_1 a_4 (\gamma-1)}{a_1 a_4 (\gamma+1) - a_2 a_3 (\gamma-1)} .
\end{aligned}
\end{equation}
\end{linenomath}
Once $\psi(r,\theta) = P(r)F(\theta)$ is determined, the in-plane velocity components are
\begin{linenomath}
\begin{equation} \label{eq:stokes11a}
\begin{aligned}
u_r(r,\theta) &= - r^\gamma \Big[ B_1 (\gamma+1) \sin\big((\gamma+1)\theta\big) - B_2 (\gamma+1) \cos\big((\gamma+1)\theta\big) \\
&\quad + B_3 (\gamma-1) \sin\big((\gamma-1)\theta\big) - B_4 (\gamma-1) \cos\big((\gamma-1)\theta\big) \Big]
\end{aligned}
\end{equation}
\end{linenomath}
and
\begin{linenomath}
\begin{equation} \label{eq:stokes11b}
\begin{aligned}
u_\theta(r,\theta) &= - (\gamma+1) r^\gamma \Big[ B_1 \cos\big((\gamma+1)\theta\big) + B_2 \sin\big((\gamma+1)\theta\big) \\
&\quad + B_3 \cos\big((\gamma-1)\theta\big) + B_4 \sin\big((\gamma-1)\theta\big) \Big].
\end{aligned}
\end{equation}
\end{linenomath}
From these relations, the Stokes solution for the velocity components along the Cartesian axes is obtained,
\begin{linenomath}
\begin{equation} \label{eq:stokes12}
v^{\mathcal{S}}(r,\theta) = u_r \sin\theta + u_\theta \cos\theta 
\quad \text{and} \quad
w^{\mathcal{S}}(r,\theta) = u_r \cos\theta - u_\theta \sin\theta.
\end{equation}
\end{linenomath}
The corresponding pressure field can be then obtained from the radial component of the Stokes equation:
\begin{linenomath}
\begin{equation} \label{eq:stokes13}
\frac{\partial p^{\mathcal{S}}}{\partial r} = \mu \Bigg( \frac{\partial^2 u_r}{\partial r^2} + \frac{1}{r}\frac{\partial u_r}{\partial r} + \frac{1}{r^2}\frac{\partial^2 u_r}{\partial \theta^2} - \frac{2}{r^2}\frac{\partial u_\theta}{\partial \theta} - \frac{u_r}{r^2} \Bigg),
\end{equation}
\end{linenomath}
which integrates to
\begin{linenomath}
\begin{equation} \label{eq:solution}
p^{\mathcal{S}}(r,\theta) = 4 \mu \gamma r^{\gamma-1} \Big[ -B_3 \sin\big((\gamma-1)\theta\big) + B_4 \cos\big((\gamma-1)\theta\big) \Big],
\end{equation}
\end{linenomath}
up to an additive constant.
At this point, all the unknowns in Equations~\eqref{eq:lambda-coco} have known expressions, which can be evaluated at every grid point prior to the DNS simulation.

It is worth discussing the choice of the exponent $\gamma$. Since we are interested in representing the Stokes solution in the immediate vicinity of sharp corners, i.e., as $r \to 0$, the dominant behavior is controlled by the smallest eigenvalue $\gamma$, while the other eigenvalue contributes only as a subdominant term.
For a flat wall, corresponding to a tip angle $\alpha = \pi$, the eigenvalues are $\gamma_1 = 1$ and $\gamma_2 = 2$, which yield velocity fields that are linear and quadratic functions of the distance from the wall, respectively, consistent with a Taylor expansion. As $r \to 0$, the linear contribution associated with $\gamma_1$ dominates.
For smaller tip angles, the two eigenvalues remain distinct, until the limit $\alpha = 0$ is reached, corresponding to a cusp. In this extreme case, both modes have the same order, $\gamma_1 = \gamma_2 = 0.5$, and it is not possible to determine a priori which mode dominates.
In principle, retaining both eigensolutions would yield a higher-order approximation of the COCO coefficients. 
However, for sharp corners with finite tip angles, as considered here, it is sufficient to retain only the dominant exponent $\gamma_1$, which is significantly smaller than $\gamma_2$. 
Accordingly, all calculations in the present work are performed for finite tip angles using only the dominant eigenvalue, $\gamma = \gamma_1$.

\begin{figure}
	\centering
	\includegraphics[width=\linewidth]{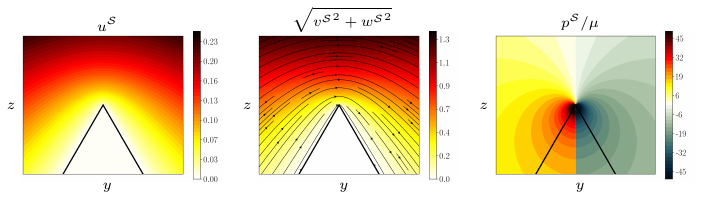}
	\caption[Solutions for the Laplace and Stokes problems]{Eigensolutions for the Laplace ($u^{\mathcal{S}}$) and Stokes ($v^{\mathcal{S}}$, $w^{\mathcal{S}}$, $p^{\mathcal{S}}$) problems. The Stokes solutions are shown for both the eigenvalue $\gamma=\gamma_1$. In the vector plots, the horizontal component is $v^{\mathcal{S}}$, the vertical one $w^{\mathcal{S}}$. The Stokes solution for pressure $p^{\mathcal{S}}$ presents a singularity at the riblet tip. The pressure plot shows the ratio $p^{\mathcal{S}} / \mu$, as the value for viscosity does not influence the corner correction coefficients values, as visible from Equations~\eqref{eq:Stk-coco}, \eqref{eq:solution}.}
	\label{fig:analyticalstokes}
\end{figure}

As an example, the eigensolutions corresponding to this dominant eigenvalue for the transverse ($v^{\mathcal{S}}$) and vertical ($w^{\mathcal{S}}$) velocity components, as well as for the pressure ($p^{\mathcal{S}}$), are shown in Figure~\ref{fig:analyticalstokes}.

A further observation must be made regarding the pressure gradient term in Equation~\eqref{eq:lambda-coco}.  
When a required velocity point in Equation~\eqref{eq:lambda-coco} lies inside the solid the corresponding velocity is set to zero. 
Unlike velocity points, pressure points that fall inside the solid require a different treatment, as the Stokes pressure field $p^{\mathcal{S}}$ is, in general, nonzero at the fluid–solid boundary (see Fig.~\ref{fig:analyticalstokes}).
The analytical prolongation of the local Stokes solution into the riblet is polydromic: continuing from the two sides of the symmetry axis yields distinct branch values below the riblet tip. 
Although one might be tempted to use, for each side, the branch continuous with that side, the discrete pressure unknown in the solver represents a single degree of freedom, which is calculated from the pressure-correction step to enforce continuity.
To remain consistent with the numerical pressure field, we prescribe a single-valued, artificial prolungation of $p^{\mathcal{S}}$ inside the solid body.
Such pressure field is continuous with the in-fluid pressure values from Equation~\eqref{eq:solution}, vanishes as $\theta = \pm \pi$, and takes intermediate values between the symmetry axis and the walls.
An example of multiple velocity points ($\mathbf{x_1}$, $\mathbf{x_2}$) sharing the same ``in-body'' pressure point is illustrated in Figure~\ref{fig:polydromy}.

\subsection{Remarks on the corner-correction coefficients}
\label{section:coco-other}

We close this section with some observations on the behaviour of the $\lambda_{\alpha}^{\mathcal{S}}$ values.  
As the Laplace/Stokes solution evaluated along the radial direction increasingly resembles a linear function with distance from the tip, the COCO coefficients vanish for large values of $r$.  
This indicates that the COCO correction has a diminishing impact on the solution far from the boundary, and that computing COCO coefficients at such locations does not improve the solution, as these coefficients are effectively zero.

\begin{figure}
	\centering
	\includegraphics[width=0.5\linewidth]{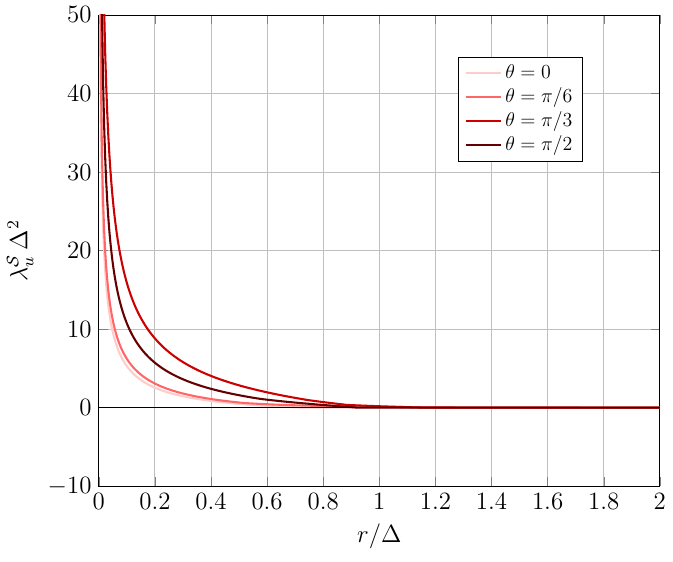}
	\caption[COCO coefficients]{Corner correction coefficient for the longitudinal velocity component rescaled by the square of the reference grid spacing in the transverse plane. The coefficients are plotted as function of the distance from the riblet tip for different values of the $\theta$ angle. It can be noticed that all coefficients seem to vanish when the distance from the tip exceeds one reference grid spacing. It can be shown that the coefficients for the other velocity components present a similar behaviour.}
	\label{fig:coeff}
\end{figure}

Consequently, a finite radius of application for the COCO correction can be defined, avoiding unnecessary calculations.  
Figure~\ref{fig:coeff} shows the values of the COCO coefficients $\lambda_x^{\mathcal{S}}$ for the longitudinal velocity component, multiplied by $\Delta^2$ (here $\Delta = \Delta_x = \Delta_y = \Delta_z$), as a function of the distance from the tip, $r/\Delta$.  
The results are shown for a riblet tip angle of $60^\circ$; however, the trends remain qualitatively similar for different tip angles and grid resolutions.

As expected, all curves decrease monotonically with increasing $r/\Delta$, and the coefficients effectively vanish beyond approximately one grid spacing from the tip.  
A similar behaviour is observed for $\lambda_y^{\mathcal{S}}$ and $\lambda_z^{\mathcal{S}}$.
In some cases the coefficient computed from Eq.~\eqref{eq:lambda-coco} may be negative due to the interplay between the Stokes solutions and the grid spacing. To preserve numerical stability, negative values are set to zero.

Based on this analysis, selecting a COCO radius of approximately twice the reference grid spacing ensures that all relevant COCO coefficients are captured.  
It is also worth noting that, similarly to the IBM, the coefficients grow rapidly as the wall is approached, enforcing the velocity to vanish.  
At larger distances from the tip, along the ribbed-wall direction, the COCO coefficients progressively converge towards the IBM values enforcing a linear velocity profile, i.e.\ $\lambda_{\alpha}^{\mathcal{S}} \approx \lambda^{\mathcal{I}}$.

Hereinafter, the COCO radius is set equal to twice the grid spacing. When the grid spacings differ between the transverse and wall-normal directions, the radius is chosen as twice the larger spacing, ensuring that all relevant contributions from the Stokes solution are included.

\section{Results}
\label{section:results}
In this section, the previously introduced corner correction is applied to two example problems.
Both involve a riblet geometry, for which this approach is particularly beneficial and for which the impact of the modification on the IBM terms can be readily assessed.

In Section~\ref{section:laminar}, the corner correction is applied to a Stokes problem, for which the enforced near-corner Stokes solution coincides with the exact one.
For a laminar flow ($Re > 0$), this property would hold only for $u$, and not for $v$ and $w$.
Here, a Stokes flow around a riblet is simulated by imposing unit far-field velocities in the longitudinal and transverse directions and setting the nonlinear terms to zero.
This setup allows the computation of the protrusion heights, i.e. two characteristic lengths of the Stokes flow in the longitudinal and transverse directions.
The resulting values are compared with the reference data of \citet{luchini-manzo-pozzi-1991}, enabling a direct comparison between the IBM and COCO implementations.

The second example, presented in Section~\ref{section:turbulent}, considers a turbulent channel flow over ribbed walls. 
The corner correction is applied at all riblet tips, and results are reported in terms of the friction difference between ribbed and smooth walls. 
Validation is performed against the literature data of \citet{endrikat-etal-2021, wong-etal-2024}, obtained using extremely fine, body-conforming grids.

\subsection{The Stokes flow over a riblet}
\label{section:laminar}

\begin{figure}
	\centering
	\includegraphics[width=0.6\linewidth]{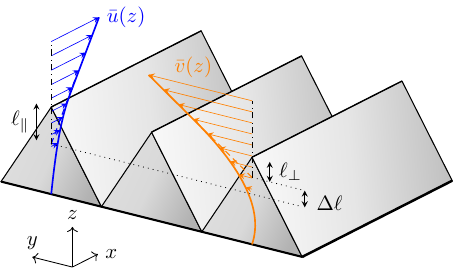}
	\caption{A visual representation of the difference between the protrusion heights, $\Delta \ell = \ell_\parallel - \ell_\perp$. The blue (orange) velocity profile represents the $y$-averaged longitudinal (transverse) Stokes flow over the ribbed wall. $\ell_\parallel$ is the longitudinal protrusion height, $\ell_\perp$ the transverse one.}
	\label{fig:ph}
\end{figure}

For riblet geometries, their behavior in the Stokes flow regime is particularly relevant, as drag reduction is purely viscous. 
Since their effect is confined to the near-wall region, the ribbed surface is typically represented using virtual origins, which denote the vertical position at which a smooth wall would reproduce the same $y$-averaged velocity profile as the ribbed wall. 
This concept can be described in terms of protrusion heights, as introduced by \citet{luchini-manzo-pozzi-1991}, which correspond to the distance between the riblet tips and the virtual origin.
In Stokes flows, riblets exhibit different protrusion heights for longitudinal and transverse directions, offering greater resistance to transverse flow. 
As a result, the virtual origin for longitudinal flow lies lower than that for transverse flow, i.e., the longitudinal protrusion height ($h_\parallel$) is larger than the transverse one ($h_\perp$).
Protrusion heights are usually normalized by the square root of the groove area, $\ell_g$ \cite{garcia-mayoral-jimenez-2011a}, making them purely geometrical parameters. 
The corresponding dimensional values are $\ell_\parallel = h_\parallel \ell_g$ and $\ell_\perp = h_\perp \ell_g$.
Figure~\ref{fig:ph} shows a representation of the protrusion heights for a riblet geometry. 

When drawing an analogy with turbulent flows, the longitudinal velocity component corresponds to the mean velocity field, while the transverse component represents turbulent spanwise fluctuations. 
Consequently, if the mean flow experiences a lower virtual origin (or equivalently a larger channel height) than turbulence, turbulence is damped, leading to drag reduction relative to a smooth surface. 
This viscous mechanism, dominant for very small riblet sizes, can be captured numerically only if the protrusion heights are accurately reproduced, as the difference between the longitudinal and transverse protrusion heights ($\Delta h = h_\parallel - h_\perp$) determines the frictional behavior of riblets at small scales \cite{bechert-etal-1997, luchini-1996, garcia-mayoral-jimenez-2011a}. 
Given their importance in turbulent flows, it is essential that protrusion heights are represented accurately at the grid resolutions typically used for turbulent DNS.

\begin{figure}
	\centering
	\includegraphics[width=0.85\linewidth]{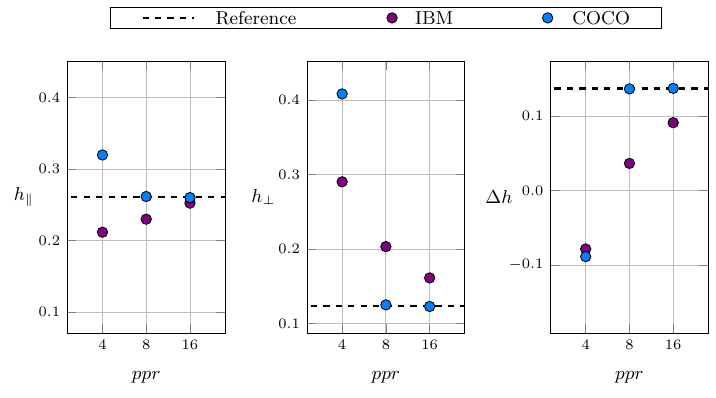}
	\caption{Computed protrusion heights values compared to the reference ones from \citet{luchini-manzo-pozzi-1991} (dashed lines). IBM indicates that the corner-correction is deactivated and the simulation is run with the immersed boundary for every point confining with the boundary. The azure points are obtained with corner-correction, using the eigenvalue $\gamma=\gamma_1$ to solve the Stokes problem.}
	\label{fig:ph_calc}
\end{figure}

Protrusion heights can be computed numerically with high accuracy for any groove shape \cite{luchini-manzo-pozzi-1991}, as shown in Fig.~\ref{fig:ph_calc}. 
Here, we reproduce the reference results of \citet{luchini-manzo-pozzi-1991} using the finite-difference code described in Section~\ref{section:method} with the corner correction activated, discretizing the domain with a moderate number of points comparable to that used in standard turbulent simulations of smooth channels.

In DNS of turbulent channel flows, sufficient wall-normal resolution typically imposes the most stringent requirement. 
However, when riblets are present on the walls, spanwise resolution becomes equally critical \citep{garcia-mayoral-jimenez-2011,modesti-etal-2021}. 
To generate a Stokes flow around the riblet geometry, we set up a Couette flow in the longitudinal and transverse directions, maintaining uniform shear far from the wall while setting the nonlinear terms to zero. 
The channel semi-height ($\delta$) is unity, measured from the half-channel to the virtual origin of the longitudinal flow, and the riblet size for this test case is $\ell_g/\delta = 0.03$. 
To assess the grid resolution required for turbulent simulations, we test different spanwise spacings of 4, 8, and 16 points per riblet (ppr), while maintaining approximately 16 points per riblet in the wall-normal direction. 
Coarser wall-normal resolutions are not considered, as DNS meshes are typically highly refined to resolve steep near-wall gradients. 
In the wall-normal direction, the spacing is uniform up to the riblet tip and then stretched toward the channel centerline; spacing far from the wall is less critical, as the velocity profile is linear in this region.

The solution accuracy is assessed based on the retrieved longitudinal and transverse protrusion heights. 
The slope of the $y$-averaged velocity profile from the Stokes simulations is used to extrapolate the virtual origin positions and the corresponding protrusion heights. 
These heights are computed both using the simple IBM of \citet{luchini-etal-2025} and the COCO approach at the riblet tips described in Section~\ref{section:coco}.
The aim is to demonstrates that, at the same grid resolution, and thus at the same computational cost, the COCO terms yield more accurate results than the IBM terms.

Results are shown in Figure~\ref{fig:ph_calc}. 
With the IBM correction, predictions for $h_\parallel$, $h_\perp$, and $\Delta h$ improve monotonically with grid resolution; however, for the present cases, the relative error in $\Delta h$ remains above 30\%. 
Achieving grid independence would therefore require finer spacing in both directions. 
The dominant contribution to the residual error comes from $h_\perp$: even on the finest grid, a 31\% error remains, whereas the error in $h_\parallel$ is only 3\%. 
When COCO is applied near the riblet tip, the protrusion heights are significantly improved. 
The $h_\parallel$ estimates match the reference values closely for 8 and 16 points per riblet (ppr), with errors below 1\%, while a notable error persists at 4 ppr. 
Similarly, COCO does not yield an acceptable $h_\perp$ at 4 ppr, but the error drops below 1\% as the resolution increases.

These results indicate that the COCO correction significantly improves accuracy: with at least 8 points per riblet in the spanwise direction (and \(\approx 16\) in the wall-normal direction), protrusion heights can be computed with errors below 1\%. 
Achieving a similar level of accuracy with the canonical IBM of \citet{luchini-etal-2025} would require a substantially higher number of degrees of freedom. 

For the case with only 4 ppr in the transverse direction, the local Stokes behavior near the corner cannot be recovered, despite using the same governing equations as those for the COCO coefficients. 
This occurs because the analytic Stokes field used to compute the forcing coefficients is an eigensolution for an infinite 2D wedge, whereas the simulations involve finite riblets with periodic $y$-boundary conditions. 
With the correction radius set to twice the maximum grid spacing (see \S~\ref{section:coco-other}), the infinite-wedge eigensolution is enforced well outside the near-tip region and, for 4 ppr, up to the periodic $y$ boundary, which degrades the results.

In Section~\ref{section:coco-vw}, we argued that selecting the Stokes eigenvalue $\gamma = \gamma_1$ and neglecting $\gamma_2$ is sufficient to represent the local Stokes solution, since the smaller eigenvalue dominates near the corner. 
The results of this section confirm this hypothesis: computing the COCO coefficients using $\gamma_1$ alone yields accurate protrusion heights with a moderate number of points. 
While including both eigenvalues could further improve accuracy, the single-eigensolution approach already provides satisfactory results. 
%

\subsection{The turbulent channel flow over ribbed walls}
\label{section:turbulent}

In this section, we assess the performance of the COCO approach in turbulent channel flows over ribbed walls. Extending the study to turbulent flows is crucial, since although inexpensive Stokes simulations can indicate potential drag-reduction benefits through protrusion heights \cite{luchini-1996}, they cannot reliably predict the total skin-friction reduction in fully turbulent channels.
In this case, imposing the Stokes solution near the riblet tip is only an approximation, as turbulence introduces significant convective and unsteady effects close to the wall. 
We nevertheless assume that, in the immediate vicinity of solid boundaries and particularly near sharp corners, these contributions remain higher-order compared to viscous diffusion. 
This is justified by the extremely large velocity gradients at geometric singularities, which make the Laplacian term dominant. 
By enforcing a locally steady Stokes solution around sharp corners, the near-wall flow can be accurately captured even with relatively coarse grids, offering an advantage over standard immersed-boundary treatments at the same resolution. 

Due to the random nature of turbulence, all flow variables are expressed using Reynolds decomposition and time averaging. 
Specifically, the fluctuating component $\phi'$ is defined as the difference between the signal $\phi$ and its average $\overline{\phi}$ over time and the periodic homogeneous directions: $\phi' = \phi - \overline{\phi}$.

Hereafter, the superscript $^{+}$ denotes quantities in wall, or viscous, units, i.e., nondimensionalized using the friction velocity $u_\tau = \sqrt{\tau_w / \rho}$ and the viscous length $\delta_\nu = \nu / u_\tau$, based on the wall shear stress $\tau_w$ and fluid viscosity $\nu$. 
We also introduce the friction Reynolds number, $Re_\tau = u_\tau \delta / \nu = \delta / \delta_\nu$, where $\delta$ is the half-channel height.

\subsubsection{The virtual origins framework to characterize turbulent flows over riblets}
\label{section:turbulent-basics}


Traditionally, the effect of a textured surface on skin friction is expressed as the relative change in the friction coefficient compared to a smooth wall, $1 - C_f / C_{f,0}$, where $C_f = 2/(U_b^+)^2$ and $U_b$ is the channel bulk velocity. The $_0$ subscript refer to the reference smooth wall case.
Since $U_b^+$ depends on the Reynolds number, metrics based on $C_f$ are inherently $Re$-dependent \cite{garcia-mayoral-jimenez-2011a, gatti-quadrio-2016}.
A more robust, $Re$-independent and now standard measure is the vertical displacement of the mean velocity profile in the logarithmic region, $\Delta U^+$, defined as
\begin{linenomath}
\begin{equation} \label{eq:dU}
\Delta U^+ = \left.\overline{u_0}^+\right|_{log} - \left.\overline{u}^+\right|_{log} \,,
\end{equation}
\end{linenomath}
where $\left.\overline{u_0}^+\right|_{log}$ and $\left.\overline{u}^+\right|_{log}$ are the averaged smooth-wall and ribbed-wall velocity profiles evaluated in the log layer. 
By convention \citep{jimenez-2004}, $\Delta U^+ > 0$ indicates a downward shift (drag increase), while $\Delta U^+ < 0$ indicates an upward shift (drag reduction).

The shift $\Delta U^+$ can be expressed as the difference between the turbulent protrusion height $\ell_T^+$ and the mean-flow protrusion height $\ell_U^+$:
\begin{linenomath}
\begin{equation} \label{eq:dU-l}
\Delta U^+ = \ell_T^+ - \ell_U^+.
\end{equation}
\end{linenomath}
These protrusion heights generalize the transverse and longitudinal protrusion heights introduced in Section~\ref{section:laminar}. 
For riblets in the viscous regime ($\ell_g^+ < 8$), the relationship $\Delta U^+ = \ell_\perp^+ - \ell_\parallel^+ = -\Delta h \ell_g^+$ holds, so that the mean- and turbulence-based protrusion heights coincide with the Stokes protrusion heights, as the riblets are fully immersed in the viscous sublayer. 
For larger riblets, turbulent simulations are required to quantify $\ell_U^+$, $\ell_T^+$, and $\Delta U^+$.

While Equation~\eqref{eq:dU} holds, it is strictly valid only if the ribbed and smooth profiles share the same turbulence virtual origin $z_T^+$.
\citet{ibrahim-etal-2021} showed that $z_T^+$ corresponds to the vertical shift (positive downward) needed to align the near-wall Reynolds shear stress profile $\overline{u'w'}^+$ of the ribbed case with that of the smooth wall.
If the ribbed and smooth-wall Reynolds stress profiles coincide, the virtual origin of turbulence is $z_T^+ = 0$, else it corresponds to the vertical offset between the two profiles.
In order to account for possible discrepancies in the $z_T^+$ positions for the ribbed and the smooth cases, the ribbed profile must be shifted by $-z_T^+$ in the wall-normal direction prior to applying Equation~\eqref{eq:dU}.
The protrusion height for turbulence represents the distance between the riblet tip and turbulence virtual origin.

The virtual origin for the mean flow can be obtained either in a Stokes-like manner, by linearly extrapolating the near-wall mean profile \citep{wong-etal-2024}, or, if $\Delta U^+$ is known, via $\ell_U^+ = \ell_T^+ - \Delta U^+$. 
Here, we adopt the latter approach. 
Care must be taken in defining viscous units in the ribbed channel: the effective channel height should account for the turbulence virtual origin, treated as the effective wall position. 
This modifies the wall shear stress and friction velocity, which becomes $u_\tau = u_{\tau,0} \sqrt{1 - z_T}$, where $u_{\tau,0}$ is the friction velocity of the reference smooth-wall case.
 
\subsubsection{Turbulent simulations set-up and grid convergence study}
\label{section:turbulent-setup}

In our simulations, the channel dimensions are $(2\pi, \pi, 2)\,\delta$ in the streamwise ($x$), spanwise ($y$), and wall-normal ($z$) directions, with riblets covering both walls. 
The domain is sufficiently large to capture outer-layer motions and first-order statistics \citep{lozano-duran-jimenez-2014a}. 
In the wall-normal direction, $z \in [\ell_\parallel - k,\, 2\delta - \ell_\parallel + k]$, where $k$ is the riblet height, $\ell_\parallel$ the dimensional longitudinal protrusion height (from literature \cite{luchini-manzo-pozzi-1991} or laminar simulations, Section~\ref{section:laminar}), and $\delta$ the channel half-height. 
This choice does not maintain a unitary average channel height, since $z=0$ is not at the riblet average height, but it provides an effective $Re_\tau$ close to the smooth-wall case. 
Placing the tip at $z = \ell_\parallel$ is preferred because $\ell_T$ (the turbulent protrusion height) is unknown a priori, and $\ell_T$ and $\ell_\parallel$ are relatively close. 
In the laminar case, this configuration returns the same friction coefficient for smooth and ribbed walls. 

The target friction Reynolds number is $Re_\tau = 400$, relatively large for riblet DNS to minimize roughness-induced blockage effects. 
This choice also matches the values used by \citet{endrikat-etal-2021,wong-etal-2024}, enabling direct comparison. 
The actual $u_\tau$ and $Re_\tau$ values are later corrected by accounting for turbulent virtual origins (Section~\ref{section:turbulent-basics}). 
Statistics are averaged over 80 time units, defined as the large-eddy turnover time $\delta / u_\tau$.

\begin{figure}
    \centering
    \begin{subfigure}[b]{0.59\textwidth}
        \centering
        \includegraphics[width=\linewidth]{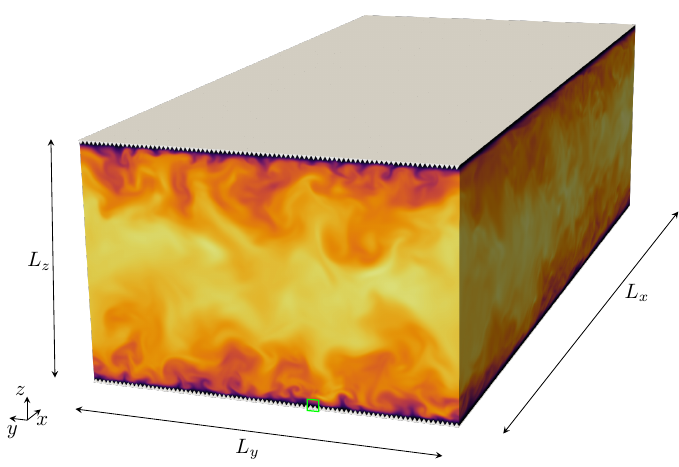}
    \end{subfigure}%
    \hfill
    \begin{subfigure}[b]{0.38\textwidth}
        \centering
        \includegraphics[width=\linewidth]{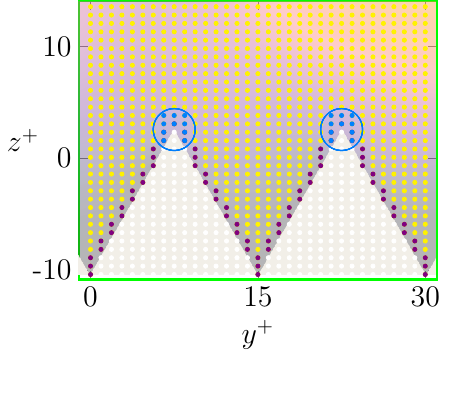}
    \end{subfigure}
    \caption{On the left, the computational domain is shown along with a snapshot of the velocity field. In this case, the riblet spacing is $15$ wall-units. The domain dimensions are reported in Table~\ref{tab:parameters}. On the right, a zoomed-in view of the flow field is presented, to focus on the grid around the wall-region. The zoomed area corresponds to the green square in the left plot. In the right plot, white dots represent grid points lying outside the fluid domain, while yellow dots indicate points inside the domain that have a null immersed boundary coefficient. Violet points are internal to the domain and adjacent to the boundary, where the classical immersed boundary coefficient is used, whereas blue points indicate regions where the corner-correction is applied. The radius of application for the corner-correction is set to $2$ times the largest grid spacing in the yz-plane.} 
    \label{fig:domain_and_zoom}
\end{figure}  
The flow is forced via a Constant Pressure Gradient (CPG), and periodic boundary conditions are enforced in the $x$ and $y$ directions. 
The riblets are represented thanks to the combination of the IBM \cite{luchini-etal-2025} and COCO (\ref{section:coco}) in the vicinity of corners. 
The numerical method used to simulate the incompressible Navier-Stokes equations is that presented in Section~\ref{section:method}. 
A visual representation of the computational domain is provided in Figure~\ref{fig:domain_and_zoom}, together with a representation of the grid for the streamwise velocity component for one of the simulated cases.
The riblets geomerty investigated in this paper is triangular, with a $60^{\circ}$ tip angle, and both walls are ribbed; the only variable geometrical parameter is the riblets size, $\ell_g$, which is adjusted to achieve different $\ell_g^+$ values.
Thanks to the IBM-COCO implementation in cartesian coordinates, which eliminates the need for a dedicated collocated grid for each geometry, the only parameters to be set are the grid spacings.
In the streamwise and wall-normal directions we select standard spacings for second-order finite-differences IBM codes, with $\Delta x^+ \approx 6$, $\Delta z^+_w \leq 0.8$ at the wall, and a natural grid stretching \cite{pirozzoli-orlandi-2021} in the wall-normal direction.
The stretching starts from the coordinate $z=0$.
The only parameter which is variable based on the considered riblet size is the spanwise spacing $\Delta y^+$; for this reason, we perform a grid convergence study to assess reasonable conditions on $\Delta y^+$. 

From Section~\ref{section:laminar}, it was evident that 4 points per riblet (ppr) in the spanwise direction are insufficient to properly resolve the Stokes flow. 
This limitation arises because the COCO coefficients are computed using the Stokes eigensolution for an infinite 2D wedge, which does not fully match the periodicity of finite riblets. 
The issue occurs only for the 4 ppr case: the correction region (radius $\approx 2 \Delta y$) spans the entire riblet spacing, while the region over which the coefficients are significant ($\approx \Delta y$, see Figure~\ref{fig:coeff}) is only a quarter of the riblet spacing, so the correction is no longer confined to the corner vicinity. 
For the turbulent simulations, we select a riblet size $\ell_g^+ \approx 10$ and test the same spanwise resolutions as in the Stokes cases: 4, 8, and 16 ppr, corresponding to spanwise spacings of 3.75, 1.88, and 0.94 wall units, respectively.
\begin{table}
    \centering
    \begin{tabular}{@{\extracolsep{8pt}} cccccccccc @{}}
        \cmidrule[\heavyrulewidth](l{-5pt}r{-5pt}){1-10}
        & $\ell_g^+$ & $s^+$ & $L_x^+$ & $L_y^+$ & $\delta ^+$ & $\Delta x^+$ & $\Delta y^+$ & $\Delta z^+$ & $ppr$ \\
        \cmidrule[\heavyrulewidth](l{-5pt}r{-5pt}){1-10}
IBM & 9.9 & 15.0 & 2516.0 & 1260.0 & 400.5 & 5.99 & 3.75 & 0.80 $\div$ 5.11 & 4.0 \\ 
IBM & 9.9 & 15.0 & 2516.0 & 1260.0 & 399.9 & 5.99 & 1.88 & 0.80 $\div$ 5.10 & 8.0 \\ 
IBM & 9.9 & 15.0 & 2750.0 & 1260.0 & 399.7 & 5.95 & 0.94 & 0.80 $\div$ 5.10 & 16.0 \\ 
COCO & 9.9 & 15.0 & 2516.0 & 1260.0 & 399.9 & 5.99 & 3.75 & 0.80 $\div$ 5.10 & 4.0 \\ 
COCO & 9.9 & 15.0 & 2516.0 & 1260.0 & 399.5 & 5.99 & 1.88 & 0.80 $\div$ 5.10 & 8.0 \\ 
COCO & 9.9 & 15.0 & 2516.0 & 1260.0 & 399.5 & 5.99 & 0.94 & 0.80 $\div$ 5.10 & 16.0 \\ 
        \cmidrule[\heavyrulewidth](l{-5pt}r{-5pt}){1-10}
    \end{tabular}
    \caption{Geometric and discretization parameters for the cartesian grids used. The terms "IBM" and "COCO" represent cases in which the simple IBM is used, and cases in which the corner-correction is activated close to the riblets tips. $\ell_g$ represents the square root of the groove cross section, $s$ the riblet tip-to-tip spacing, $L_x$, $L_y$ are the domain dimensions in the periodic directions, $\delta$ represents the channel half height. $\Delta x$, $\Delta y$, $\Delta z$ represent the grid resolutions in the three directions, while $ppr$ is the number of points per riblet in the spanwise direction.}
    \label{tab:parameters-gc}
\end{table}
While the 4 ppr case is expected to underperform, comparing the 8 and 16 ppr cases is instructive. 
For turbulence simulations, we rely on the assumption that, near the corner, nonlinear and unsteady effects are higher order relative to viscous effects. 
Consequently, the corner-correction radius should remain small, not exceeding the extent of the viscous sublayer. 
Since the COCO coefficients are significant over a radius $\approx \Delta y$ and the viscous sublayer is typically $\sim 5$ wall units thick, all cases are reasonable in this regard. 
The grid convergence study is also performed with the standard IBM \citep{luchini-etal-2025} using the same grid parameters. 
In the wall-normal direction, the minimum spacing is kept fixed, as in Section~\ref{section:laminar}, to match the smooth-wall $\Delta z_w^+$ requirement at the wall. We achieve grid convergence without reducing $\Delta z_w^+$. 
Increasing $\Delta z_w^+$ would be inappropriate, as turbulence would not be resolved \citep{pope-2000}. 
Simulation parameters for this study, including riblet geometry, domain size, and grid spacings, are summarized in Table~\ref{tab:parameters-gc}.
\begin{figure}
	\centering
	\includegraphics[width=0.45\linewidth]{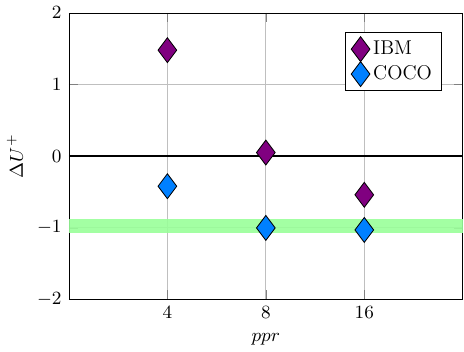}
	\caption{$\Delta U^+$ values obtained for the triangular riblet geometry at $\ell_g^+ \approx 10$ as a function of the number of points per riblet ($ppr$) in the spanwise direction. Violet markers represent cases simulated with the standard IBM from \citet{luchini-etal-2025}, the blue ones are simulated with the corner-correction. The green shaded area represents the $\Delta U^+$ value reported by \citet{wong-etal-2024} accounting for a $95\%$ confidence interval on the measure, obtained for the same cross-section geoemetry, with $\ell_g^+ = 9.7$.}
	\label{fig:dUppr}
\end{figure}

Figure~\ref{fig:dUppr} shows the results of the grid convergence study, presenting the shift in the mean velocity profile, $\Delta U^+$, as a function of points per riblet (ppr) in the spanwise direction.  
For the canonical IBM, the convergence towards the COCO values is slow: even at the finest grid, $\Delta U^+$ differs by $\approx 0.5$ wall units, indicating that much finer grids, would be needed to match the corner-correction results. This is consistent with the Stokes simulations, where the 16 ppr IBM case gave $\Delta h \approx 30\%$ off the reference \citep{luchini-manzo-pozzi-1991} and the COCO results.  

For the COCO-activated simulations, as expected, the 4 ppr case fails to produce acceptable results, while the 8 and 16 ppr cases yield nearly identical $\Delta U^+$ values, differing by only $\approx 0.03$ wall units, or $\sim 2.7\%$ of the measure. This difference is smaller than the temporal uncertainty, $1.96\,\sigma_{\Delta U^+} \approx 0.06$ wall units at a 95\% confidence level, where $\sigma_{\Delta U^+}$ is the standard deviation of $\Delta U^+$. The latter was estimated via Gaussian error propagation from the standard deviations of $U_b^+$ and $U_{b,0}^+$, assuming a logarithmic mean velocity profile ($\Delta U^+ \approx (U_{b,0}^+ - U_b^+)/\delta^+$). Individual $U_b$ statistics were computed from the time series using an unbiased batch means method with adaptive batch size \cite{russo-luchini-2017}.  

From this study, we conclude that, when the corner-correction is applied, 8 ppr in the spanwise direction are sufficient to achieve accurate results.

\subsubsection{Turbulent results and discussion}
\label{section:turbulent-setup}

Having established a suitable resolution for accurately discretizing turbulent flow over riblet geometries, we perform parametric simulations of triangular riblets with a $60^\circ$ tip angle to obtain their drag-reduction curve and validate it against literature data.  
Turbulent channel flows are simulated at $Re_\tau = 400$ for six riblet sizes, covering the relevant portion of the drag-reducing regime ($\ell_g^+ \approx 4-17$).  
For three of these cases, simulations are also performed using the canonical IBM with the same grid as the COCO cases.  
All numerical parameters are listed in Table~\ref{tab:parameters}, and a minimum of 8 points per riblet in the spanwise direction is used in all simulations.
\begin{table}
    \centering
    \begin{tabular}{@{\extracolsep{8pt}} cccccccccc @{}}
        \cmidrule[\heavyrulewidth](l{-5pt}r{-5pt}){1-10}
        & $\ell_g^+$ & $s^+$ & $L_x^+$ & $L_y^+$ & $\delta ^+$ & $\Delta x^+$ & $\Delta y^+$ & $\Delta z^+$ & $ppr$ \\
        \cmidrule[\heavyrulewidth](l{-5pt}r{-5pt}){1-10}
IBM & 5.9 & 9.0 & 2516.0 & 1260.0 & 399.8 & 5.99 & 0.64 & 0.80 $\div$ 5.11 & 14.0 \\
IBM & 9.9 & 15.0 & 2750.0 & 1260.0 & 399.7 & 5.95 & 0.94 & 0.80 $\div$ 5.10 & 16.0 \\
IBM & 14.1 & 21.5 & 2750.0 & 1268.5 & 399.8 & 5.95 & 1.34 & 0.80 $\div$ 5.11 & 16.0 \\
COCO & 3.9 & 6.0 & 2516.0 & 1260.0 & 399.7 & 5.99 & 0.60 & 0.80 $\div$ 5.10 & 10.0 \\
COCO & 5.9 & 9.0 & 2516.0 & 1260.0 & 399.6 & 5.99 & 0.64 & 0.80 $\div$ 5.11 & 14.0 \\
COCO & 7.9 & 12.0 & 2516.0 & 1260.0 & 399.5 & 5.99 & 0.86 & 0.80 $\div$ 5.10 & 14.0 \\
COCO & 9.9 & 15.0 & 2516.0 & 1260.0 & 399.5 & 5.99 & 0.94 & 0.80 $\div$ 5.10 & 16.0 \\
COCO & 11.8 & 18.0 & 2516.0 & 1260.0 & 399.5 & 5.99 & 1.00 & 0.80 $\div$ 5.11 & 18.0 \\
COCO & 14.1 & 21.5 & 2516.0 & 1268.5 & 399.5 & 5.99 & 1.34 & 0.80 $\div$ 5.11 & 16.0 \\
COCO & 17.1 & 26.0 & 2516.0 & 1274.0 & 399.7 & 5.99 & 1.62 & 0.79 $\div$ 5.08 & 16.0 \\
        \cmidrule[\heavyrulewidth](l{-5pt}r{-5pt}){1-10}
    \end{tabular}
    \caption{Geometric and discretization parameters for the cartesian grids used. The terms "IBM" and "COCO" represent cases in which the simple IBM is used, and cases in which the corner-correction is activated close to the riblets tips. $\ell_g$ represents the square root of the groove cross section, $s$ the riblet tip-to-tip spacing, $L_x$, $L_y$ are the domain dimensions in the periodic directions, $\delta$ represents the channel half height. $\Delta x$, $\Delta y$, $\Delta z$ represent the grid resolutions in the three directions, while $ppr$ is the number of points per riblet in the spanwise direction.}
    \label{tab:parameters}
\end{table}

\begin{figure}
	\centering
	\includegraphics[width=0.85\linewidth]{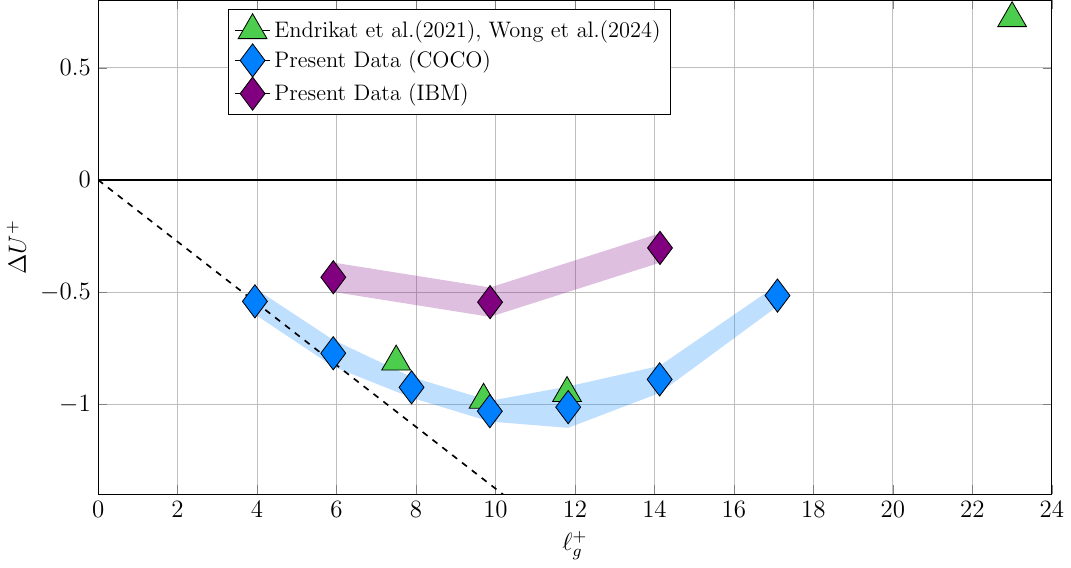}
	\caption{$\Delta U^+$ values obtained as a funciton of $\ell_g^+$, with the shaded bands representing the temporal uncertainty of the measure for a $95 \%$ confidence interval. The dashed line represents the viscous regime, defined by $\Delta U^+ = -\Delta h \, \ell_g^+$. The plus-adimensionalization is performed considering the value of $\tau$ at the origin of turbulence $z=z_T$ as $\tau_{w}$. Triangular markers correspond to DNS results from \citet{endrikat-etal-2021,wong-etal-2024}. As regards the two data points coming from \citet{endrikat-etal-2021}, the $\Delta U^+$ values from the most recent evaluation \cite{wong-etal-2024} were selected.}
	\label{fig:dU0}
\end{figure}
Figure~\ref{fig:dU0} presents the roughness function $\Delta U^+$ for all riblet sizes, with and without corner correction.  
When using COCO, our results reproduce the drag-reduction curve with high fidelity, showing good agreement with the DNS data of \citet{endrikat-etal-2021,wong-etal-2024}, obtained in minimal-channel simulations.  
The datasets of \citet{endrikat-etal-2021} and \citet{wong-etal-2024} predict slightly lower drag reduction compared to our results; nevertheless, the overall trends are consistent, with the maximum drag reduction for the present geometry reaching approximately $\Delta U^+ = -1.0$.
The dashed line, $\Delta U^+ = - \Delta h \ell_g^+$, represents the viscous regime. Previous studies \cite{jimenez-1994, bechert-etal-1997} suggested that the slope in this regime should be reduced by a factor $\mu_0$ ($0.66$ and $0.785$, respectively), which we attribute to experimental limitations such as finite tip radius, manufacturing imperfections, or insufficient spatial resolution that prevent an accurate representation of the near-corner flow.  
When riblets are fully immersed in the viscous sublayer, their effect on the flow should match the prediction from the corresponding Stokes flow. This is confirmed by the smallest simulated geometries, whose $\Delta U^+$ values lie on the dashed line, consistent with \citet{luchini-1996}.  
The maximum drag reduction occurs for $\ell_g^+ \approx 10-12$, in agreement with literature \cite{garcia-mayoral-jimenez-2011a}, while $\Delta U^+$ increases for larger riblet sizes, indicating the onset of the drag-increasing regime around $\ell_g^+ \approx 20$.  
Consistent with observations from the laminar simulations (Figure~\ref{fig:ph_calc}) and the grid-convergence study (Figure~\ref{fig:dUppr}), the impact of the corner correction is evident: simulations without COCO predict roughly half the drag reduction of the corner-corrected cases at the same grid resolution.  
Shaded areas indicate the $95\%$ confidence interval for $\Delta U^+$, with an average standard deviation of about $0.03$ wall units.

All results, in terms of the roughness function $\Delta U^+$, the percentage change in friction coefficient ($\Delta C_f \% = (1-C_f/C_{f,0}) \cdot 100$), and the associated standard deviations, are summarized in Table~\ref{tab:results}.  
The trends in $\Delta C_f \%$ closely follow those of $\Delta U^+$, since the simulations are performed in a full-channel configuration, which allows all turbulent scales to be correctly represented, unlike minimal-channel setups \cite{chung-etal-2015, endrikat-etal-2021}.  
Standard deviations for $\Delta C_f \%$ were computed analogously to those for $\Delta U^+$, using the temporal histories of $C_f$ for both the smooth and ribbed cases.  
On average, the uncertainty in the $\Delta C_f \%$ estimates is approximately $0.15\%$.

\begin{table}
    \centering
    \begin{tabular}{@{\extracolsep{16pt}} ccccccccc @{}}
        \cmidrule[\heavyrulewidth](l{-9pt}r{-9pt}){1-9}
         & $\ell_g^+$ & $s^+$ & $ppr$ & $t \, \left(\delta / u_\tau\right)$ & $\Delta C_f \%$ & $\sigma_{\Delta C_f \%}$ & $\Delta U^+$ & $\sigma_{\Delta U^+}$ \\
        \cmidrule[\heavyrulewidth](l{-9pt}r{-9pt}){1-9}
	  IBM & 5.9 & 9.0 & 14.0 & 79.9 & -4.46 & 0.17 & -0.43 & 0.03 \\
	  IBM & 9.9 & 15.0 & 4.0 & 128.2 & 18.10 & 0.17 & 1.48 & 0.03 \\
	  IBM & 9.9 & 15.0 & 8.0 & 111.7 & 0.54 & 0.17 & 0.05 & 0.03 \\
	  IBM & 9.9 & 15.0 & 16.0 & 98.3 & -5.43 & 0.17 & -0.54 & 0.03 \\
	  IBM & 14.1 & 21.5 & 16.0 & 114.1 & -2.93 & 0.17 & -0.30 & 0.03 \\
	  COCO & 4.0 & 6.0 & 10.0 & 95.7 & -5.61 & 0.15 & -0.54 & 0.03 \\
	  COCO & 5.9 & 9.0 & 14.0 & 100.6 & -7.68 & 0.15 & -0.77 & 0.03 \\
	  COCO & 7.9 & 12.0 & 14.0 & 118.4 & -9.00 & 0.12 & -0.92 & 0.02 \\
        COCO & 9.9 & 15.0 & 4.0 & 124.8 & -4.34 & 0.17 & -0.42 & 0.03 \\
        COCO & 9.9 & 15.0 & 8.0 & 110.4 & -9.63 & 0.17 & -1.00 & 0.03 \\
        COCO & 9.9 & 15.0 & 16.0 & 118.4 & -9.83 & 0.17 & -1.03 & 0.03 \\
	  COCO & 11.8 & 18.0 & 18.0 & 127.5 & -9.69 & 0.23 & -1.01 & 0.05 \\
	  COCO & 14.1 & 21.5 & 16.0 & 99.9 & -8.78 & 0.15 & -0.89 & 0.03 \\
	  COCO & 17.1 & 26.0 & 16.0 & 108.6 & -5.12 & 0.13 & -0.52 & 0.02 \\
        \cmidrule[\heavyrulewidth](l{-9pt}r{-9pt}){1-9}
    \end{tabular}
    \caption{A summary of all simulated geometries and the respective drag reduction $\Delta C_f \%$ and shift in mean velocity profile $\Delta U^+$. $\ell_g^+$ is the square root of the groove cross section and $s^+$ is the riblets spacing in plus-units. $\sigma_{\Delta C_f \%}$ and $\sigma_{\Delta U^+}$ represent the standard deviations for the drag reduction and $\Delta U^+$ values due to the finite simulation time $t$. Data from the grid convergence study was also included in this table.}
    \label{tab:results}
\end{table}

\begin{figure}
	\begin{subfigure}{.48\textwidth}
		\centering
		\includegraphics[width=\linewidth]{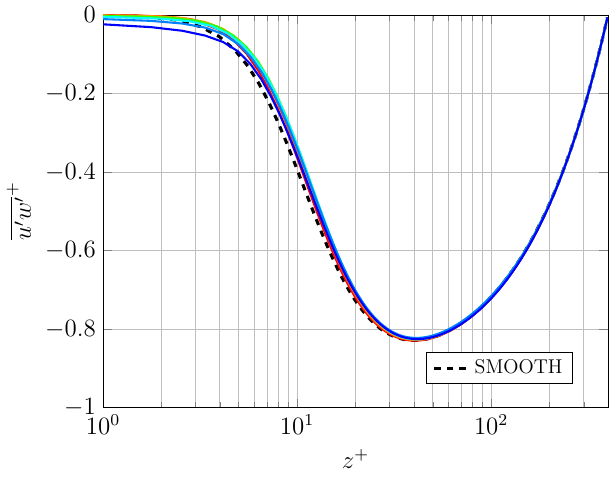}
		\label{fig:uw_in}
	\end{subfigure}
	\hfill
	\begin{subfigure}{.48\textwidth}
		\centering
		\includegraphics[width=\linewidth]{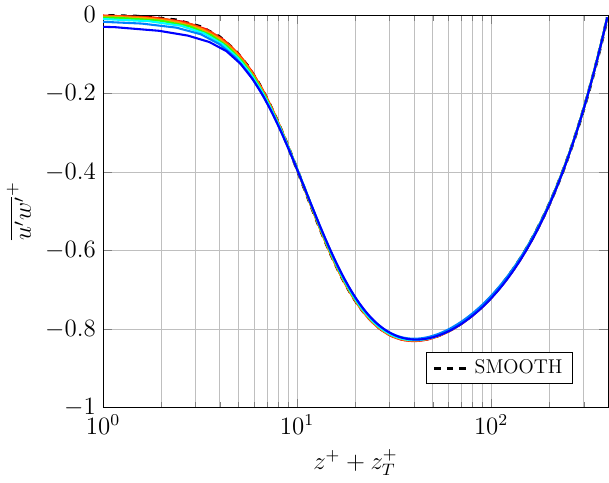}
		\label{fig:uw_fin}
	\end{subfigure}
	\caption{Turbulent shear stress $\overline{u^\prime w^\prime}^+$ profiles as functions of the wall distance $z^+$. The dashed line represents the reference smooth channel, while the colored lines are the ribbed channels. On the left the profiles from the simulations are plotted. On the right the ribbed profiles are shifted by $z_T$ and rescaled with $u_\tau = u_{\tau,0} (1-z_T)^{0.5}$, to account for the turbulence virtual origin. The color scale goes from red ($\ell_g^+ \approx 4$) to blue ($\ell_g^+ \approx 17$). Only profiles with the corner-correction activated are shown. }
	\label{fig:uw}
\end{figure}

\begin{figure}
	\begin{subfigure}{.48\textwidth}
		\centering
		\includegraphics[width=\linewidth]{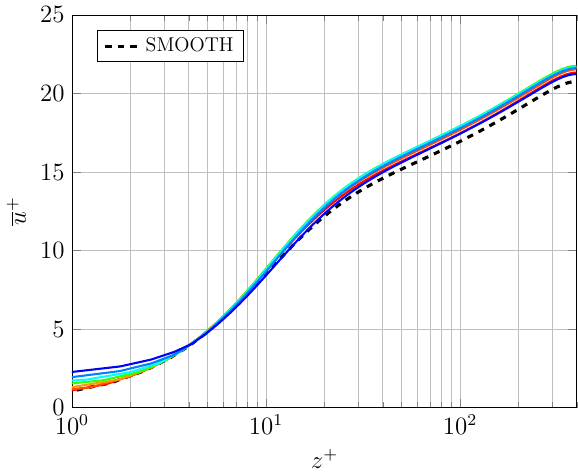}
		\label{fig:U_in}
	\end{subfigure}
	\hfill
	\begin{subfigure}{.48\textwidth}
		\centering
		\includegraphics[width=\linewidth]{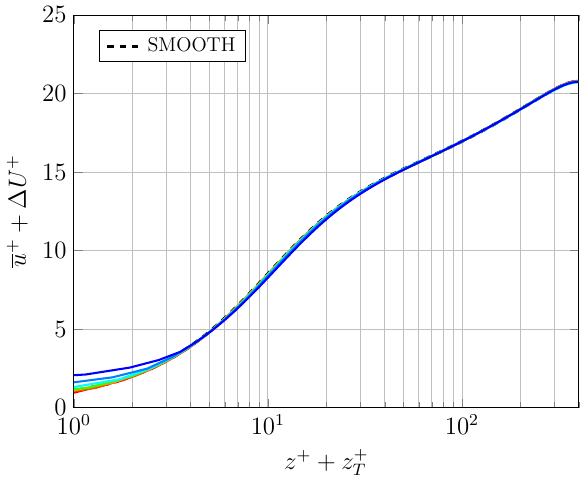}
		\label{fig:U_res}
	\end{subfigure}
	\caption{The left plot represents mean velocity profiles coming from simulation output. The right plot presents them shifted in the wall-normal direction by $z_T^+$, and in the vertical direction by $\Delta U^+$, which is found by overlapping the ribbed mean velocity profiles on the smooth one in the logaritmic region. On the right plot all profiles are rescaled with the $\tau_w$ value accounting for the turbulence virtual origin.}
	\label{fig:U}
\end{figure}

To illustrate the smooth-wall-like behaviour of turbulence over ribbed surfaces, as discussed in \citet{ibrahim-etal-2021}, Figures \ref{fig:uw} and \ref{fig:U} show that, once the turbulence virtual origin $z_T^+$ and the mean-velocity shift $\Delta U^+$ are accounted for, the Reynolds shear stress and mean velocity profiles for ribbed and smooth walls collapse onto each other.  
In each figure, the left panels display the raw (uncorrected) simulation results, while the right panels show the profiles shifted to account for the virtual origin.  
Additionally, in the right panels the profiles are rescaled using the appropriate friction velocity $u_\tau$, as described in Section~\ref{section:turbulent-basics}.

%

\begin{figure}
	\begin{subfigure}{.49\textwidth}
		\centering
		\includegraphics[width=\linewidth]{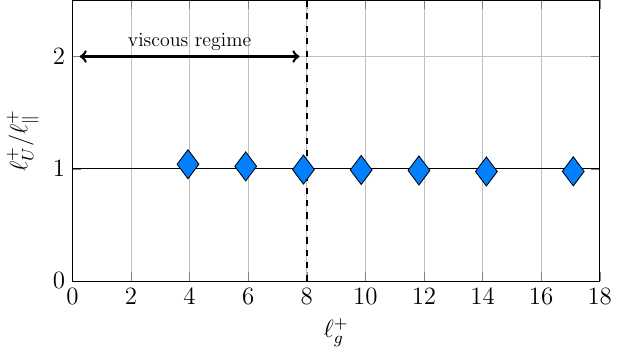}
		\label{fig:ell_X}
	\end{subfigure}
	\hfill
	\begin{subfigure}{.49\textwidth}
		\centering
		\includegraphics[width=\linewidth]{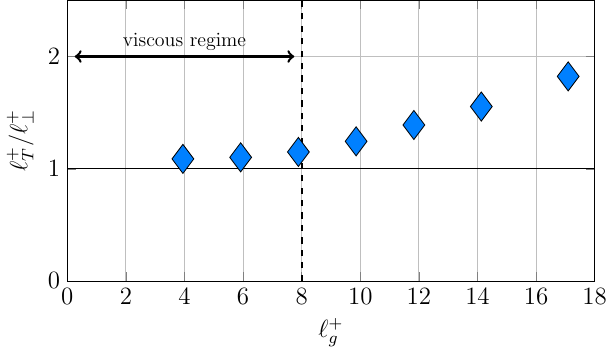}
		\label{fig:ell_Y}
	\end{subfigure}
	\caption{Mean flow ($\ell_U^+$) and turbulent ($\ell_T^+$) protrusion heights are compared to the longitudinal ($\ell_X^+$) and transverse ($\ell_Y^+$) ones for all simulated riblets sizes ($\ell_g^+$).}
	\label{fig:ell}
\end{figure}

As discussed above, using the values of $\ell_T^+$ and $\Delta U^+$ obtained from the $\overline{u}^+$ and $\overline{u^\prime w^\prime}^+$ profiles, the mean-flow protrusion height can be computed as $\ell_U^+ = \ell_T^+ - \Delta U^+$.  
In the viscous regime, typically for $\ell_g^+ < 8$, the protrusion heights for the mean and turbulent flows are expected to closely match the longitudinal and transverse protrusion heights.
For larger riblet sizes, however, the turbulence protrusion height deviates from the transverse one, as a virtual origin for the spanwise flow alone no longer suffices to accurately capture the turbulence origin, due to the growing contribution of higher-order wall-normal velocity fluctuations \citep{ibrahim-etal-2021}.  
Figure~\ref{fig:ell} shows that $\ell_U^+$ scales almost perfectly with $\ell_X^+$ across all $\ell_g^+$ values, while $\ell_T^+$ follows $\ell_Y^+$ for the smallest riblets but diverges for larger $\ell_g^+$.

We conclude this section with a remark on simulation costs.  
A comparison can be made with the results of \citet{endrikat-etal-2021}, despite differences in domain configuration and solver.  
A domain-independent and practically relevant metric is the cost per riblet at fixed riblet size and Reynolds number: we consider the geometry with a $60^\circ$ tip angle at $\ell_g^+ \approx 10$ and $Re_\tau \approx 400$.  
As reported in \citet{endrikat-etal-2021} for their case T615, the streamwise and wall-normal resolutions are comparable to those adopted here, so that the dominant differences in computational cost arise from the timestep and spanwise point density.  
Our implementation allows a timestep of $\Delta t^+ \approx 0.10$ (in wall units) versus $\Delta t^+ \approx 0.03$ reported in \citet{endrikat-etal-2021}.  
The larger timestep is primarily due to our Cartesian mesh with corner-correction, which does not require grid refinement at the riblet tips, making stability constraints less restrictive than for the body-fitted, tip-refined grid used in the reference.  
In the spanwise direction we use $16$ points per riblet (ppr) versus $29$ ppr in the reference.  
Treating computational cost as proportional to (points)$\times$(timesteps) gives the approximate ratio $(16/29)\times(0.03/0.10) \approx 0.17$, indicating that our implementation is roughly six times less expensive.  
This estimate is conservative, as we assume the per-point, per-timestep cost is equal for both solvers, although a simple Cartesian finite-difference solver is expected to be more efficient than a body-fitted finite-volume code.  
Furthermore, mesh-convergence tests confirm that $8$ ppr in the spanwise direction are sufficient for our solver to obtain accurate results, further enhancing the cost efficiency of our methodology.
In terms of total computational cost, a full-channel simulation with $\ell_g^+ = 14$ and active corner correction, covering $\approx 100$ large-eddy turnover times and using $420 \times 944 \times 355 = 140,750,400$ degrees of freedom, required $\approx 14,600$ CPU-core hours.  
Simulations were performed on the Hawk HLRS cluster using $1536$ cores distributed over $14$ nodes, each equipped with two AMD EPYC 7742 CPUs.

\section{Conclusions}
\label{section:conclusions}

In this paper, we have presented an innovative approach for resolving flows near sharp corners.  
The underlying idea is that viscous effects dominate in the immediate vicinity of sharp edges, where velocity gradients are largest.  
Enforcing the local Stokes solution in these regions successfully improves the accuracy of standard immersed-boundary methods (IBM), which typically approximate the wall with a linear fit and require substantially finer grids to achieve comparable results, even relative to classical body-fitted meshes.  
The proposed method, termed corner-correction (COCO), builds upon the immersed-boundary approach of \citet{luchini-etal-2025}.  
It modifies the IBM coefficients near geometric singularities using expressions based on the local Stokes solution, which is available analytically for sharp corners, and on the grid spacing.  
This allows accurate representation of the near-corner flow with relatively coarse grids.

The corner-correction was evaluated for two distinct test cases: a Stokes flow, where the local Stokes solution perfectly represents the actual flow, and a turbulent channel flow, where the Stokes solution is only valid in the immediate vicinity of sharp corners.  
Both cases employ ribbed walls, chosen because they represent a particularly challenging scenario for accurately resolving near-corner flows.  
In the Stokes cases, the method accurately reproduced the protrusion heights using very few points per riblet, outperforming the standard IBM on the same grid.  
For turbulent simulations, the same spanwise resolution per riblet as in the Stokes tests was sufficient to capture the characteristic drag-reduction curve of the riblets.  
Once again, the corner-correction proved essential for achieving accurate results at reduced computational cost: a full DNS at $Re_\tau = 400$ required approximately 15,000 CPU-core hours.  
This efficiency is enabled by a smooth-wall-like grid, with a spanwise spacing of roughly 1 viscous unit and a wall-normal spacing of $0.8$ viscous units at the wall.

For the turbulent cases, our methodology accurately reproduced the roughness function, capturing the linear viscous regime, the drag-reduction peak, which agrees well with literature values, and the friction increase for larger $\ell_g^+$.  
When accounting for the virtual origin of the smooth-wall cases, the turbulent Reynolds shear stress and mean velocity profiles collapse onto the corresponding smooth-channel data, confirming that drag-reducing riblets preserve a smooth-wall-like turbulence structure.  
By comparing the turbulent protrusion heights with the Stokes counterparts, we found that the mean-flow protrusion height follows the same trend as the longitudinal Stokes protrusion, whereas the turbulent protrusion height deviates from the transverse Stokes protrusion for $\ell_g^+ \geq 8$.

The versatility of this implementation allows for straightforward extension to any geometry featuring sharp corners, such as irregular wall surfaces or bluff bodies, as long as a Stokes solution is available.  
For alternative riblet designs, the only parameter that needs adjustment is the tip angle.  
In cases where analytical solutions are unavailable, numerical Stokes solutions can be used to compute the forcing coefficients without compromising accuracy or computational efficiency.  
For moving boundaries, the forcing coefficients must be updated at each timestep, yet the framework remains robust.  
Future research will exploit this framework to investigate innovative wall 
geometries aimed at maximizing drag reduction and enhancing heat transfer, 
while also addressing open questions regarding the physics of turbulent flows 
over ribbed and more general structured surfaces.

\section*{Data availability}
Data will be made available on request.

\section*{Declaration of competing interest}
The authors declare that there are no conflicts of interest.

\section*{Acknowledgments}
The project was partially funded by the Deutsche Forschungsgemeinschaft (DFG, German Research Foundation), grant number 521110788.
The simulations were performed on the HPE Apollo (Hawk) supercomputer at the High Performance Computing Center Stuttgart (HLRS) under grant number 44290.
The authors gratefully acknowledge Federica Gattere, Giorgio Maria Cavallazzi, Giacomo Ronchetti, and Andrea Rossi for their contributions during the early stages of this project.


\printcredits

\bibliographystyle{cas-model2-names}
\bibliography{Wallturb}

\end{document}